**Title**:
Ultra-high frequency ultrasound imaging and quantification of microvascular flow in xenograft renal cell carcinoma in an avian chorioallantoic membrane model

**Running Title:**
UHFUS flow imaging and quantification in tumors in CAM

**Authors**:
Sara Mar[1] MSc, Emmanuel Chérin[2] PhD, Justin Xu[2] BASc, David E. Goertz[1,2] PhD, Hon S. Leong[1,2] PhD, Christine E.M. Demore[1,2] PhD

**Affiliations**:

1. Department of Medical Biophysics, University of Toronto, Toronto, Ontario, Canada
2. Sunnybrook Research Institute, Toronto, Ontario, Canada

**Corresponding author**:

Emmanuel Chérin
(emmanuel.cherin@sri.utoronto.ca)
Sunnybrook Research Institute
Physical Sciences (Room S638)
2075 Bayview Avenue
Toronto, Ontario, M4N 3M5, Canada
Phone: 416-480-6100, ext. 63277

**Footnote:**
Sara Mar is a Precision Medicine Associate at AstraZeneca Ltd, Mississauga, ON, Canada.
Justin Xu is a PhD candidate at the Big Data Institute, University of Oxford, UK.

**Abstract**:

Patient derived xenograft (PDX) tumor models initiated in avian chorioallantoic membranes (CAM) are under investigation to evaluate the effectiveness of therapeutic options with the objective of personalizing treatments. CAM PDXs paired with ultra-high frequency ultrasound (UHFUS) imaging could potentially constitute prospective high throughput assays that can rapidly assess tumor volume and vascular response to therapy. To date, little work has been conducted to adapt and validate UHFUS flow imaging methods to CAM tumor models. Here we report the development and evaluation of an imaging pipeline for UHFUS detection of microvascular flow in a CAM tumor model using interframe subtraction (IS) to suppress tissue clutter. The IS pipeline included a tissue motion compensation (MC) stage prior to clutter filtering and was compared to a singular value decomposition (SVD) clutter filter. The performance was evaluated using UHFUS data acquired in phantom and in vivo Sunitinib-treated renal cell carcinoma.

MC substantially reduced tissue motion effects. MC+IS was comparable to MC+SVD filtering at detecting flow within tumors. The results for both IS and SVD filters were dependent on the details of implementation. The UHFUS imaging methods detected a significant decrease in blood flow metrics in treated versus control tumors.

An effective imaging pipeline was developed for the assessment of the treatment response of CAM PDX models in a clinically relevant timeframe. The MC+IS approach implemented on B-scan image derived data is less computationally intensive and can be used with widely available UHFUS systems.

# I. Introduction

Avian hosts are an emerging alternative to mouse models for xenografts such as patient derived xenografts (PDXs) for drug sensitivity testing, and whose responses to multiple potential therapies could be assessed at early stages of treatment. Avian tumor models are accessible, cost effective, and can be scaled for applications such as drug sensitivity assays [1,2]. This model involves grafting tumors onto the chorioallantoic membrane (CAM), otherwise known as the "embryonic lung", of either a chicken or duck embryo. The CAM is a highly vascularized and rapidly growing microvascular tissue bed in which engrafted xenografts vascularize within 1-2 days post implantation, and relevant information on tumor response can be obtained in as little as 10 days [1–4].

Detecting and measuring changes in tumor volume and vasculature is important for assessment of tumor response to therapy in pre-clinical models. Vascular targeting agents such as anti-angiogenic therapies, which disrupt vascular proliferation pathways [5–7], have been used to treat highly angiogenic cancers such as metastatic renal cell carcinoma (mRCC), and more recently, combined with immune targeting therapies to have more potent anti-tumor effects [8–10]. However, anti-angiogenics and combination therapies still suffer from low response rates and drug resistance, leading to multi-line treatments for many patients, particularly those with metastatic disease [7,11–14]. Highly angiogenic cancers therefore represent prime examples of diseases that would benefit from PDX tumor models. For this to be achieved, a cost effective and accessible *in vivo* model and imaging method must be implemented to assess the treatment response in a clinically relevant time frame.

Imaging using ultra-high frequency ultrasound (UHFUS) systems offers a non-invasive, rapid, cost effective and contrast-free method to assess longitudinal changes in tumor vasculature and volume [15–17]. UHFUS is suitable for imaging avian PDXs since the vascular bed is superficial [18–22], and major preclinical imaging facilities across the world offer access to high-frequency ultrasound systems. These systems typically use conventional power Doppler flow imaging, which is ineffective in detecting smaller microvessels due to the combined effects of tissue motion, high-pass clutter filtering, and the low signals levels associated with slowly moving microvascular blood flow [23–25]. However, using a commercial preclinical system (Vevo 3100 and MX 700 probe, FUJIFILM-VisualSonics Inc., Toronto, Canada), Huang *et al.* [24] recently developed an ultrasound microvessel imaging (UMI) method using quadrature demodulated (IQ) data collected in power Doppler mode. A singular value decomposition (SVD) based clutter filter was applied to interleaved B-mode IQ data acquired in that mode to isolate the blood signal, which was then used to generate microvessel power Doppler images. From these images, two metrics quantifying tumor vessel density and perfusion were derived. They demonstrated superior detection of the tumor microvasculature and higher sensitivity to the response to anti-angiogenic therapies with UMI compared to conventional power Doppler imaging acquired with the same system. A number of limitations could affect the widespread adoption of UMI [24], such as using IQ data, which are not necessarily available on all UHFUS systems. In addition, SVD filtering can require a large ensemble of IQ frames for proper suppression of clutter signal and therefore be computationally demanding. Defining and automating SVD thresholding to produce consistent microvasculature images with minimal biases also presents a challenge [26]. Furthermore, without tissue motion compensation, cyclic cardiac motion and random embryo body movement exhibited by the CAM could negatively impact UMI performance [27].

To address some of these limitations, we propose in this study a UHFUS imaging method that includes (i) tissue motion compensation, (ii) clutter filtering using interframe subtraction (IS) and (iii) speckle temporal variance as a metric for blood flow and/or blood volume. This method was developed using data collected *in vitro* from a vessel flow phantom and *in vivo* from CAMs and RCC (Renca) tumors grown in CAMs, using a Vevo 2100 ultrasound imaging system (FUJIFILM-VisualSonics Inc.) equipped with a MS 700 (50 MHz) probe. Although all processing steps involved in our UHFUS method only require uncompressed signal envelopes, which could be extracted from RAW cineloop frames, IQ data were collected for benchmarking against UMI, and uncompressed envelopes calculated from it. Compensation for tissue motion was implemented to account for gross tissue motion due to cardiac activity and/or embryo movement. The sensitivity of speckle variance to flow was investigated in flow phantom experiments for a range of velocities and in temperature-controlled flow experiments in an RCC tumor. IS clutter filtering was evaluated in experiments in the flow phantom, a CAM vascular bed, and 12 RCC tumors, for a range of interframe times (IT). The effect of tissue motion on speckle variance was also investigated in these tumors. Finally, our UHFUS imaging method was used to evaluate the response of RCC tumor vasculature to an anti-angiogenic therapy (sunitinib) [3,4,10,24,28], with results compared to fluorescent immunohistochemistry findings.

# II. Materials and Methods

In the following sections, the establishment of the CAM tumor model and histological assessment of tumor perfusion is described. Then, the UHFUS imaging workflow and setup used for CAM imaging is presented, followed by the signal processing methods, including tissue motion compensation and clutter filtering. The experimental timeline for UHFUS method validation and antiangiogenic treatment assay is depicted in Figure 1.

## A. CAM Tumor Model

Renca cells were obtained from Western University (London, Ontario, Canada) and grown in culture medium (RPMI 1640, Wisent, Quebec, Canada) supplemented with 10% fetal bovine serum. Cells were incubated at 37°C and 5% $CO_2$. When reaching 80-90% confluency, cells were divided at a 1:3 ratio. Cells were resuspended from fourteen 70-90% confluent 15 cm diameter tissue culture dishes using 5 mM of ethylenediaminetetraacetic acid (EDTA) in phosphate-buffered saline (PBS). Cell suspensions were transferred to 50 mL conical tubes and centrifuged for 5 minutes at 200 g. The supernatant was removed, and the cell pellet was resuspended with PBS and centrifuged again. This washing process was performed three times in total. A Cytation 5 Cell Imaging Multimode Reader (Agilent Technologies, Santa Clara, CA) was used to count the number of cells in suspension. Renca cells (44 million) were divided into eleven 1-mL tubes. Matrigel (15 µL) was added to the cell pellets (~30 µL) and lightly mixed using a micropipette.

Fertilized duck eggs (Khaki Campbell and White Layer Hybrid ducks; Weber Farm, Ontario, Canada) were placed in a rotating humidified incubator for 4 days (Maru Deluxe, Rcom, Wichita, KS, USA) at 38°C and 45% humidity. On the fourth day of incubation, or embryonic day (ED) 4, the eggs were sterilized with 70% ethanol, lightly cracked on a sterile circular petri dish, and quickly opened to transfer the embryo, yolk sack and albumin to a 10 cm x 10 cm weigh boat. The *ex-ovo* embryos were stored in housing tubs (6-12 per tub) and transferred to a fridge incubator at 38°C and 20% humidity.

On ED 8, a 1 cm filter paper disk (Whatman grade 1, Global Life Sciences Solutions Canada ULC Vancouver, British Columbia, Canada) was placed onto a desired engraftment site on the CAM, at a midway point between the embryo body and the edge of the membrane, and quickly removed to produce a slight wound. 17.5 µL of the cell and Matrigel mixture was immediately engrafted onto the wound. 100 mL of water was added to the bottom of the housing tubs to maintain humidity. Twelve CAMs, each engrafted with $2.5 \times 10^6$ Renca cells, were used for evaluating the UHFUS imaging system, and 22 CAMs each engrafted with $2 \times 10^6$ Renca cells were used for the sunitinib treatment assay. Between imaging sessions, specimens were incubated up to ED 16, then perfused with stain for fluorescent imaging and tumors harvested for histology.

## B. UHFUS Data Acquisition

UHFUS imaging was performed using a Vevo 2100 imaging system (FUJIFILM-VisualSonics Inc., Toronto, Ontario, Canada) equipped with a 50-MHz linear array transducer (MS700) attached to a linear stepper motor (P/N 11484, FUJIFILM-VisualSonics Inc.) (Figure 2). Coupling gel (~ 0.5 mL) was placed in a plastic wrap, which was then affixed to the transmitting end of the transducer, carefully avoiding the formation of air bubbles (Figure 2). The transducer was positioned over the area of interest of the CAM, which was covered by very small amount of sterile PBS to facilitate the translation of the transducer over its surface and maintain acoustic coupling between probe and specimen (Figure 2 C).

The system was operated in B-mode for real-time display, with the system's RF-mode activated to enable the collection of IQ data. For blood flow imaging, 630 frames of beamformed B-mode IQ data were acquired at each of three planes in the sample, one at the center and two at equidistant locations on either side (usually 1-3 mm from center). Imaging was performed with settings of "100%" transmit power, "high" line density, "standard" sensitivity, a single transmit focal zone, and a 1-8 mm (depth) x 9.37 mm (width) field of view. These parameters were selected to maximize the frame rate (107 fps). For this field of view, I and Q were each 584 samples x 512 lines in size (per frame), resulting in a pixel size of approximately 12 μm axially by 18 μm laterally. The position of the transmit focal depth (7.5 mm) was chosen such that the amplitude of the signal acquired from a tissue mimicking phantom (4% w/w $SiO_2$ particles, S5631-500G, Sigma-Aldrich, mixed in 2% w/w agar, #281210, BD Difco Agar Technical) was as close to uniformity as possible over the imaging depth (up to 8 mm) required for tumor imaging. No time gain compensation was applied to avoid depth variation of the noise level.

For tumor volume assessment, a linear stepper motor was used to acquire sequential B-mode images of the tumors at a 0.1 mm step size. The settings for B-mode image acquisition were identical to those used for flow images except three transmit focal zones (set at 4.5 mm, 5.25 mm, and 6 mm) were used, and the 'sensitivity' was set to high. Tumors were manually contoured in all acquired B-mode images using VevoLab (v. 5.7.1, FUJIFILM-VisualSonics) to estimate their volume.

## C. UHFUS Data Processing

Our proposed UHFUS data processing method is summarized in Figure 3 (green path). Following data acquisition and export of beamformed in-phase (I) and quadrature (Q) signals, the uncompressed signal envelope (E) is calculated as $E = \sqrt{(I^2 + Q^2)}$. Alternatively, it can be reconstituted from standard logarithmically compressed RAW data [29] exported from the Vevo system as $E = 10^{RAW/20}$. E is used for tissue motion estimation in each of the imaging planes, and its subsequent compensation. Then, an IS clutter filter is applied to remove tissue signal and isolate blood flow. Following IS filtering and removing frames with out-of-plane tissue motion, speckle temporal variance is calculated in each pixel across all remaining filtered frames and averaged within a manually selected region of interest (ROI). For comparison purposes, UMI was also implemented using SVD clutter filtering on in-plane tissue motion compensated data followed by calculation of the "power Doppler" signal (see below for details) in each pixel and averaging over the same ROI.

### *1. Tissue Motion Compensation*

Tissue motion might comprise either one or both in-plane and out-of-plane components. In the present study, non-rigid frame registration was used to compensate in-plane motion. The displacement matrix of a given acquired frame was estimated relative to the first frame using '*imregdemons*', a diffusion-like regularization method implemented in MATLAB (The Mathworks Inc.. Natick, MA, USA). This was done using the uncompressed signal envelope $E$ calculated from IQ data, 3 'pyramid' levels (i.e. spatial resolution levels), 100 iterations for the lowest resolution level, 1 iteration for the two higher resolution levels, and an accumulated field smoothing parameter of 3.0. These parameter values were selected because they minimized in a small group of visually stationary tumors the impact of the motion algorithm, which was quantified by the difference in mean speckle variance post interframe subtraction clutter filtering (both described below) between motion compensated and non-motion compensated data. The estimated displacement was then compensated for using '*imwarp*' (with cubic interpolation), an image warping function also provided in MATLAB. This warping, which was applied directly to the IQ data corresponding to that frame so that SVD clutter filtering could eventually be performed with in-plane tissue motion compensated data, was followed by an envelope calculation prior to IS clutter filtering. Out-of-plane tissue motion was delt with post IS clutter-filtering by dropping a subset of clutter filtered frames, based on the correlation coefficients between their constitutive in-plane motion-compensated frames and the first frame of the set, as described below in *Speckle Temporal Variance*.

### 2. *Interframe Subtraction*

Inspired from past work by our group [30] , an 'interframe' clutter filtering approach was developed, based on the subtraction of frames acquired at the same imaging plane location in the CAM tumor at different times. In the case of stationary tissue, this operation removes tissue signals, and the remaining signals are associated solely with moving blood (Eq. 1**,** Figure 4). Effectively this is a first-order finite impulse response (FIR) filter that is applied in the interframe time direction at each location within the image plane. Needles et al. [30], investigated this on IQ data collected at relatively low frame rates (0.25 and 20 fps) using a swept-scan single-element 40-MHz transducer. The filtering circumvented the effects of tissue spectral broadening induced by the motion of the transducer by effectively eliminating tissue clutter prior to in-frame Doppler processing and, consequently, enabling detection of slower flow. The present study uses array-based acquisitions and is not subject to tissue spectral broadening associated with mechanical scanning. In this case the motivation is to employ it as a clutter filter that, due to its application on an interframe time scale, has a frequency 'roll-off' that can enable the retention of very low blood velocity signals. Here, interframe subtraction was applied to the uncompressed envelope frames calculated from IQ data collected in B-mode at a higher frame rate ($FR = 107\ Hz$). This frame rate allowed us to investigate the effects of the interframe time ($IT = m/FR$, with $1 \leq m \leq 25, m \in \mathbb{N}$) on slow flow detection.

The IS clutter filter is described in Eq. 1, where $x_n$ and $x_{n+m}$ are frames collected at times $t$ and $t + IT$, respectively, while $y_n$ is the resulting clutter filtered frame resulting from pixel-wise subtraction. It is a high pass filter whose transfer function and cut-off frequency- which determine the lower blood velocity detection limit- are expressed in Eq. 2 and Eq. 3, respectively.

$$y_n = x_n - x_{n+m} \qquad \text{Eq. 1}$$

$$H_m(f) = 1 - \exp\left(j2\pi \frac{mf}{FR}\right), \text{ with } f \text{ the frequency} \qquad \text{Eq. 2}$$

$$f_m(L) = \frac{FR}{2\pi m} \cos^{-1}(1 - 2 \times L^2), \text{ with } L\ (0 \leq L \leq 1) \text{ the cut-off level} \qquad \text{Eq. 3}$$

Slower flow induces a slower decorrelation of the envelope signal [31] (in other words the envelope signal varies more slowly), which results in envelope variations in a lower frequency range. As the interframe number *m* increases, and subsequently the cut-off frequency of the IS filter decreases (Eq. 3), the filter output frequency content expends toward lower frequency components, allowing the detection of slower blood flow.

### 3. *Speckle Temporal Variance*

The speckle temporal variance (referred to as speckle variance, SV, from here on) was estimated over slow time, post IS clutter filtering, in each pixel of the field of view:

$$SV = \frac{1}{N-1} \sum_{n=1}^{N} (y_n - \bar{y})^2 \qquad \text{Eq. 4}$$

with $\bar{y}$ the average over $N$ clutter filtered frames $y$, where $N = 630 - m$ maximum. Only the IS clutter filtered frames $y_n$ calculated using in-plane motion-compensated frame pairs $(x_n, x_{n+m})$ whose coefficients of correlation with the first frame were both above 0.9, were used in Eq. 4. Therefore, $N$ is reduced by the number of clutter-filtered frames removed due to this correlation threshold step. Since SV is a quadratic function of E, a compression of $10log_{10}(SV)$ was used for speckle variance images, as opposed to $20log_{10}(E)$ for B-mode images. A region of interest (ROI) corresponding to the tumor was selected in acquired B-mode images, carefully avoiding large feeding vessels on the periphery, where highest flow occurs, and areas with low speckle brightness that were interpreted as shadowed or non-viable. The mean of the speckle variance (MSV) was then calculated in each ROI and used in this study as a metric of blood volume and/or perfusion.

#### 4. *UMI processing*

The principles of singular value decomposition involved in UMI for tissue clutter filtering can be found in Demené *et al.* [32]. In brief, this method allows the separation of highly spatially and temporally coherent signal (tissue signal, typically) from weakly coherent signal from moving scatterers (red blood cells or microbubbles). In the present study, the three-dimensional complex IQ data matrix corresponding to the 630 acquired entire frames were first reshaped into a 2D space-time Casorati matrix, which was then decomposed into three matrices containing the right and left singular vectors and a diagonal matrix of singular values (ordered from the highest to the lowest). Lower rank singular values are associated with (quasi-) stationary tissue clutter, whereas higher rank singular values can be associated with moving blood. For clutter filtering, singular values of rank below a threshold selected automatically (as described below) were set to zero, and the filtered 2D Casorati and 3D complex IQ data matrix were then sequentially reconstructed. The automatic rank threshold was determined as the rank of the singular value that is the most perpendicularly distant from the straight line joining the first and last singular values. Filtered IQ data were then used to calculate power Doppler (PD) as:

$$PD = \frac{1}{N}\sum_{n=1}^{N}(I_n^2 + Q_n^2) \qquad \text{Eq. 5}$$

with $N$ equal to the total number of frames acquired (630), whether prior in-plane motion compensation was used or not.

## D. UHFUS Flow Detection Evaluation

The UHFUS data acquisition and processing pipeline (Figure 3) was tested both *in vitro* in a flow phantom, and *in vivo* in the CAM model with and without tumors. The effects of parameters including IT and flow velocity on speckle variance, flow detection and microvessel visualization were investigated.

#### 1. *Interframe Time and Flow Velocity*

The relationship between interframe time (for IS clutter filtering), mean flow velocity and speckle variance was investigated *in vitro* using a flow phantom with a 1.5 mm diameter channel embedded an agar matrix (1% w/w, #281210, Difco Agar Technical) containing silicon dioxide particles (4% w/w, S5631-500G, Sigma-Aldrich), and the transducer aligned parallel or perpendicular to the channel. Blood mimicking fluid (BMF: 769DF, CIRS, Norfolk, VA) was circulated through the wall-less channel using a syringe pump (Standard PHD ULTRA™ CP; Harvard Apparatus, Holliston, MA) at mean flow velocities ranging from 0.1 to 15 mm/s. For each flow setting, 630 frames were acquired ($FR = 107\ fps$) in planes respectively parallel and perpendicular to the flow direction and clutter filtered using either IS or SVD (for comparison). IS filtering was performed using interframe numbers $m$ ranging from 1 to 25 (by step of 1), corresponding to interframe times ranging from ~9.35 ($1/FR$) to ~233.65 ms. The speckle variance was calculated in all pixels of the imaging plane. Finally, the mean spectral variance (MSV) was estimated in a ROI delineated inside the channel boundary on the original B-mode and plotted as a function of IT and mean flow velocity. For UMI, SVD filtering was performed using automatic thresholding as previously mentioned, and power Doppler calculated. The mean power Doppler (MPD) was then estimated in the same ROI, as a function of mean flow velocity. No tissue motion compensation was used for the phantom experiments.

The same investigation was performed *in vivo* in one CAM tumor at ED 14, in which blood flow was reduced by gradually decreasing the embryo body temperature from 36.5°C down to 6°C [33,34]. This was achieved by placing the weigh boat on ice, while a thermometer was positioned in the albumin for temperature monitoring. At each of the 36.5°C, 16°C, 11.5°C, 8.5°C, 7.5°C and 6°C temperature levels, 630 frames were acquired successively at three planes of the tumor. Images were processed with tissue motion compensation this time, and either IS or SVD filtering. Finally, tumor MSV and MPD were estimated. IS filtering was performed with ITs in the same range as in the phantom experiments.

#### 2. UHFUS Detection of individual Microvessels in CAM

To provide an illustrative example of the capability of our UHFUS imaging method to detect individual microvessels of various diameters, speckle variance images were compared with brightfield optical images of a CAM vessel bed (without tumor) acquired at ED 14. Brightfield images (top view) were first acquired using an upright microscope Nikon (SMZ18, Nikon, Tokyo, Japan) equipped with 1x lens. UHFUS B-modes (cross-section) were then acquired from the same area and processed to generate speckle variance and power Doppler images. These images were produced post tissue motion compensation and using respectively IS and SVD clutter filtering to assess their effects on the detection of various sized vessels. Distance measurement tools (NIS-Elements Basic Research, Nikon) were used to measure vessel diameters and distances between vessels observed in brightfield images, which were then compared to those observed on UHFUS flow images.

#### 3. Effect of Tissue motion Compensation and Interframe Time in vivo

A total of 36 UHFUS datasets collected from 12 CAM tumors at ED 16 were used to investigate the effects of motion compensation and IT in IS clutter filtering on CAM tumor microvasculature detection via speckle variance. Data were processed with and without motion compensation, and either IS or SVD filtering. MSV or MPD were then measured in ROIs corresponding to the tumors. An automatic IT selection was also implemented based on the variation of MSV with IT, similar to the automatic thresholding used in SVD filtering. The selected IT corresponds to the interframe time at which the MSV value is the most perpendicularly distant from the line joining the MSV values at the shortest and longest IT for the dataset (Supplementary Figure 1).

### E. UHFUS Imaging for Anti-Angiogenic Treatment Assay

Finally, our UHFUS imaging method was used to evaluate the vascular response to sunitinib, a tyrosine kinase inhibitor, in RCC (Renca) grown in the CAM model.

#### 1. Treatment protocol

Twenty-two CAM tumors were randomly divided into control and treatment groups after three days of tumor growth (ED 11). Sunitinib (LC Laboratories, Woburn, MA, USA) was dissolved in 100% dimethyl sulfoxide (DMSO). A working concentration of 10 μM of sunitinib was prepared and aliquoted into 100 μL volumes. 100% DMSO was used as the vehicle control. A daily treatment of 10 μL sunitinib (10 μM; treatment group) or of 10 μL of 100% DMSO (control group) was topically delivered using a micropipette from ED 11 to ED 15. One control CAM died between ED 8 and ED 11, and 2 treatment CAMs died prior to endpoint imaging (1 between ED 11 and ED 14, and 1 more just before ED 16).

Each CAM was placed in the UHFUS set up and B-mode images were acquired across the tumor for volume estimation. Then, 630 frames of B-mode IQ data were acquired at three planes of the tumor. An example of acquired B-mode is shown in Supplementary Video 1. The orientation of the tumor was noted on the weigh boat containing the CAM to facilitate the histological sectioning along the direction of the UHFUS imaging plane. Images were processed with tissue motion compensation, and either IS filtering (IS filtering being performed at specific and automatically selected IT values) or SVD filtering. MSV and MPD were then calculated and each averaged over the three planes in each tumor.

#### 2. *Fluorescent Immunohistochemistry and Histology*

Dylight 649 labelled *Lens Culinaris Agglutinin* (Vector Laboratories Inc., Newark, CA, USA), a lectin which binds the intraluminal surface of vessel endothelial walls [35] was diluted 1:10 with PBS prior to injection. After UHFUS imaging at ED 16, 100 µL of this diluted lectin was injected, under stereomicroscope, into an embryonic vein using a custom-made glass microneedle. The embryo was placed in the incubator for 20 minutes to allow the antibody to circulate and bind to endothelial cells. The tumor was then surgically removed, immediately fixed in 10% formalin for 24 hours and transferred to 70% ethanol. Tumors were placed in multi compartment cassettes (4-5 tumors each), processed, and embedded in paraffin wax. The paraffin blocks were sectioned into 4 µm thick slices in the same orientation as they were imaged with US. Sections were rehydrated and stained with Hematoxylin and Eosin (H&E) for tissue morphological analysis.

An upright Eclipse Ni confocal microscope (Nikon Instruments Inc., Melville, NY, USA) equipped with a 10x Plan Fluor objective (Nikon) was used for fluorescent imaging. 'Vessel wall' imaging was performed using a 'Cy5' filter, a pinhole of 4.0 Airy unit, a high voltage ('HV') of 120 V (which sets the photomultiplier tube gain) and laser power ('Laser') of 2%. Manual stitching was used to acquire whole section images. Images were exported (as *.nd2 files) and analyzed using a custom MATLAB script. A region of interest (ROI) encompassing the entire tumor area (excluding feeding vessels on the periphery) was manually selected. Vessel segmentation was performed within the ROI by automatic detection of ridges [36] followed by a series of morphological operations to suppress unconnected single pixels and holes, and skeletonization. The skeleton was then convolved with a 3x3 kernel (pixel size ~2.5 µm), assuming an average microvessel diameter of 10 µm. The relative fluorescent area (FA) was then calculated from the ratio of the number of 'vessel' pixels (i.e. of non-zero value) to the total number of pixels in the tumor ROI, including potential necrotic area:

$$FA = 100 \times \frac{\#\ of\ vessel\ pixels}{\#\ of\ tumour\ pixels} \qquad \text{Eq. 6}$$

The mean fluorescence intensity (MFI) was calculated in the same ROI as well. Both FA and MFI were then averaged over several histological sections in each tumor.

#### 3. *Statistical Analysis*

The significance of the difference between data processing methods was determined using a paired t-test whereas the significance of the differences between groups of the treatment assay was evaluated using an unpaired t-test (both with $\alpha = 0.05$). Correlations with fluorescence immunochemistry findings were investigated using Pearson correlation coefficient *r*.

# III. Results

## A. UHFUS Method Validation

#### 1. *Interframe Time and Flow Velocity*

The detection of a blood mimicking fluid flowing in a 1 mm vessel phantom was evaluated using speckle variance for a range of mean velocities from 0.1 to 15 mm/s. Examples of log-compressed speckle variance and power Doppler cross-section images of the vessel obtained after IS and SVD clutter filtering, respectively, are shown in Figure 5 A-I. A typical B-mode frame of the vessel phantom is shown in Figure 5 J to highlight the low contrast (-6 dB) between BMF in the vessel and the surrounding agar matrix. An increase of the contrast between BMF and agar matrix with mean velocity is clearly observable on cross-sections obtained following IS clutter filtering. These variations are depicted in more detail in Figure 5 K which shows, for IS clutter filtering, a quasi-linear increase of the normalized MSV in the low velocity range toward a plateau. At higher IT, a steeper slope is observed, and the plateau is reached at slower velocities. Similar results were obtained when the imaging plane lateral direction was aligned with the channel in the flow direction (data not shown). The plateau is reached when the product $flow\ velocity\ \times IT$ surpasses the IS clutter filtered signal correlation length along the flow direction. From Figure 5 L, this correlation length (here along the elevation direction) would be of the order of 100 to 200 µm, in the ROI, in our experimental conditions. On the other hand, the mean power Doppler measured post SVD clutter filtering seems to be relatively independent from flow velocity in the range of velocities investigated.

The coherence length depends on the ultrasound beam spatial variations (amplitude and/or phase), which are anisotropic for the linear ultrasound array transducer used in this study. Consequently, the random orientation of the microvasculature in CAM tumors, combined with its size and resulting blood flow distributions, should complexify speckle variance (and MSV) variations with IT. Nonetheless, the changes in speckle variance with velocity in a CAM Renca tumor, in which blood flow was presumably reduced by constriction of arterioles induced by body-cooling, seem consistent with flow vessel phantom results (Supplementary Figure 2). As temperature, and therefore flow velocity, decreases, speckle variance decreases (A to C), even in the largest CAM vessels. However, the change in flow velocity is not the only contributor to the decrease in speckle variance. Moving blood volume, which affects blood signal amplitude [37], and therefore, both speckle variance and power Doppler, is a confounding variable. Indeed, power Doppler (post SVD filtering), which was shown otherwise to be relatively independent of flow velocity in phantom experiments (~0.5dB variation across the velocity range), also decreased with body temperature in this experiment.

### 2. *UHFUS Detection of individual Microvessels in CAM*

A brightfield image and UHFUS images are compared in Figure 6, to assess the UHFUS detection of individual microvessels. Log-compressed speckle variance and power Doppler images obtained post IS (IT=18.7 ms and 74.8 ms) and SVD clutter filtering, respectively, depict vessels within the CAM that correspond to vessels in the brightfield image. Vessels larger than 150 µm (1, 5) in cross section are visible post IS clutter filtering at both ITs, as well as post SVD filtering. Smaller vessels, with diameters between 60-80 µm (2, 3, 4) are more easily distinguishable with IS filtering at a higher IT of 74.8 ms, due to higher speckle variance values.

### 3. *Effect of Interframe Time and Tissue Motion Compensation in vivo*

The effects of IT selection and tissue motion compensation on the detection of flow *in vivo* were investigated in the 12 non-treated tumors used for system evaluation. Speckle variance images of a tumor obtained with and without motion compensation and IS clutter filtering at various interframe times are shown in Figure 7. Greater and more 'diffused' intratumoral signal is observed at higher IT (Figure 7 C vs A, and G vs E), due to improved detection of slower flow in smaller vessels, as predicted by vessel phantom experiments (Figure 5 A-B vs D-E, Figure 5 K). As expected, tissue motion increased speckle variance (Figure 7 top vs middle row, Figure 7 I), and blurred speckle variance images. These effects, which are associated with the addition of the tissue velocity components to blood velocity components are somewhat more prevalent at higher IT, as the cutoff frequency of the high pass IS filter decreases with interframe time (Eq. 4, with $IT = m/FR$), leading to a higher speckle variance at any velocity in the 'linear' range (Figure 5 L) corresponding to slower flow.

The effectiveness of the tissue motion compensation algorithm is demonstrated in Supplementary Video 2, which shows, for both non motion compensated and motion compensated data, speckle variance images calculated over 16-frame moving ensembles overlaid on top of the B-mode cineloop. Tumor motion has been greatly reduced, as indicated by the increase in the average correlation coefficient of the ensembles with the first frame of the cineloops and the reduction of speckle variance associated with tumor motion. The flashes observed in speckle variance at times of significant tumor motion in the non-compensated data set are also suppressed in the motion-compensated data. Note that frame rejection was not applied in this case, to be able to follow the variations of the average correlation coefficient, which is calculated using the full frames, over the entire cineloop.

The effect of tissue motion compensation on MSV and MPD is summarized for this tumor data set in Figure 7 J. Due to the variability in tissue motion amongst samples, IS clutter filtering was completed using the individual automatically selected IT. A statistically significant difference in MSV (*t* (*degree of freedom* = 35) = 5.861; $p = 1.177\times10^{-6}$) was found between motion-compensated (*Mean* = 26366.3; *Standard Deviation* = 9173.1) and non-motion-compensated (*M* = 46390.2; *SD* = 25562.5) IS clutter filtered images. However, in a subset of tumors that were visually stationary prior to motion compensation, no statistical difference in MSV (*t(5)* = 1.670; *p* = 0.156) was found between pre- (*M* = 28421.8; *SD* = 9903.4) and post-motion compensation (*M* = 28188.1; *SD* = 9882.2), indicating that, overall, the motion compensation algorithm was not significantly affecting the signal envelope variations associated with blood flow itself. This is further supported by investigation of the effects of motion compensation on moving blood signal in a visually stationary tumor, shown in Figure 8. Three 60 µm x 54 µm ROIs were selected in 'no flow', 'medium flow' and 'fast flow' areas of the tumor (Figure 8 A), and the envelope signal decorrelation observed in those ROIs after motion compensation compared to those obtained without motion compensation (Figure 8 B). No increase in decorrelation time was observed between motion compensated and non-motion compensated signals, indicating that the MC algorithm implemented in this study did not suppress local blood motion relative to tissue.

Visually, tissue motion compensation, including frame rejection post IS filtering, reduced the intratumoral speckle variance in motion-compensated images (Figure 7 E-H), relative to non-motion-compensated speckle variance images (Figure 7 A-D). Tissue motion compensation did not significantly affect MPD (*t* (35) = -1.450; *p* = 0.156) measured in SVD clutter filtered images (no MC: *M* = 34361.9, *SD* = 13792.0; MC: *M* = 35324.3, *SD* = 13206.4). However, with motion compensation, an improvement in resolution and sharpness was observed in power Doppler images of tumors originally affected by visually apparent tissue motion (Figure 7 H vs D). In this group, the correlation coefficient between MSV and MPD (measured post IS and SVD clutter filtering, respectively) increased from 0.63 without motion compensation to 0.95 when motion compensation was implemented.

The resolutions of speckle variance images calculated pre- and post-motion compensation are compared to that of B-mode in Supplementary Figure 3. This was done in the visually stationary tumor shown in Supplementary Video 1, in a region of the image where a single CAM vessel could clearly be isolated on the B-mode image, due to the low tissue background level, and flickering speckle associated with blood flow visualized. The comparison of cross-sections of the vessel in B-mode and SV images (panel C) indicate that their resolutions are similar, in both the axial and lateral directions.

## B. UHFUS Imaging in Anti-Angiogenic Treatment Assay

Example speckle variance and power Doppler images of control and treated Renca tumors at ED 16 are shown in Figure 9; the corresponding histology images are shown in Supplementary Figure 4. It should be noted that these images are not intended to be representative images of the control and treated groups, due to the high degree of intragroup variability with respect to tumor growth, vascularity, and response to treatment. However, they do highlight the consistent similarity between speckle variance images obtained post IS clutter filtering and power Doppler images obtained through the UMI processing path, both post motion compensation. Indeed, a correlation coefficient of 0.984 between MSV and MPD measurements overall (pooled EDs and pooled control and treated tumors) was measured in this group.

The overall results of this assay are shown in Figure 10. A statistically significant decrease in MSV was observed for treated (*t* (9) = -2.482; *p* = 0.035) and control (*t* (9) = -3.806; *p* = 0.004) groups between ED 11 and ED 14, and for the treated group (*t* (8) = -4.891; *p* = 0.001) between ED 14 and ED 16 (Figure 10 A). On the other hand, there was no significant difference in MSV in the control group between ED 14 and ED 16 (*t* (9) = 0.874; *p* = 0.405). Also, whereas a statistically significant difference in MSV was found between control and treated groups at ED 16 for both a fixed IT of 74.8 ms ($t$ (14) = 3.651; $p$ = $2.62\times10^{-3}$) and an auto selected IT ($t$ (17) = 3.379; $p$ = $4.94\times10^{-3}$) (Figure 10 B), this was not the case at earlier EDs. The same trend was observed in MPD results (Supplementary Figure 5), with no significant difference for control tumors between ED14 and ED16, but a statistically significant difference in MPD at endpoint between control and treated tumors ($t$ (15) = 3.497; $p$ = $3.25\times10^{-3}$). The decrease over time of tumor MSV may be due to a progressive reduction in microvascular density during tumor growth. One must note that treatment had no effect on the evolution of tumor volume overtime, since at each ED, no statistical difference in volume was found between treated and control groups (Figure 10 C). Although the lower MSV and MPD in the treated group relative to the control group at end point is most likely related to a lower overall blood volume, possibly associated to a lower vascular density in that group, no difference in fluorescent area percentage and mean fluorescent intensity was found between the two groups Figure 10 D and Figure 10 E, respectively). In fact, no correlation was found between MSV and FA ($r \sim 0.11$, $p \sim 0.67$) or MFI ($r \sim 0.07$, $p \sim 0.79$) nor between MPD and FA ($r \sim 0.02$, $p \sim 0.95$) or MFI ($r \sim -0.008$, $p \sim 0.99$).

# IV. Discussion

In this work we presented a UHFUS imaging method to detect and quantify metrics of microvascular flow in a CAM tumor model that is compatible for use with widely available UHFUS systems. Data processing involves tissue motion compensation, and interframe subtraction clutter filtering applied to uncompressed envelope data, which could be obtained from exported RAW data from Vevo imaging systems, followed by pixel-wise speckle temporal variance estimation. This method was evaluated both *in vitro* in a flow phantom and *in vivo* in CAM and CAM tumors, and compared to UMI [24], which is based on singular value decomposition clutter filtering of IQ data and pixel-wise power Doppler calculation. Both MSV following IS filtering and UMI enable visualization and characterization of slow microvascular blood flow, which is not possible with conventional Power Doppler [23–25]. It was confirmed *in vitro* that power Doppler, calculated post SVD clutter filtering, is relatively independent of flow velocity (Figure 5 K, black curve), and is therefore primarily a measure of relative blood volume, as reported in Huang et al. [24]. On the other hand, it was shown that the speckle variance calculation, measured post IS clutter filtering, varies with blood velocity in a range determined by the interframe time (Figure 5 L). As such, it is sensitive to changes in both blood volume and velocity. Visually, SVD clutter filtered images, which map blood volume, are relatively brighter than IS clutter filtered images, predominantly in pixels presenting slower blood flow because IS images map blood volume modulated by flow velocity. This is illustrated *in vitro* in Figure 5 where the intensity in the vessel in the images generated post-IS clutter filtering increases with flow velocity, whereas it stays constant in the images generated post-SVD clutter filtering (as well as *in vivo*, in Figures 6 and 8, in small vessels). Finally, this imaging and quantification method was used to investigate the effects of sunitinib on CAM tumors in a longitudinal assay. Results indicate a vascular response 5 days after the treatment start (Figure 10 A-B), which was confirmed by UMI but not detected by our fluorescent immunohistochemistry analysis (Figure 10 D-E).

With regards to clutter filtering, IS performed with a single interframe time is computationally more efficient than SVD, as the number of operations required per filtered frame is equal to the number of pixels in a frame for IS versus $O(number\ of\ pixels\ \times number\ of\ frames)$ for SVD. Furthermore, since IS processing requires only two input frames at a time, it could be implemented for real-time imaging at the memory cost of a circular buffer holding ($m$+1) frames, or 2 successive frames if the desired interframe time is 1/FR, albeit with a more time-efficient motion compensation algorithm. On the other hand, SVD performed on a large ensemble size requires a significant amount of memory and GPUs to achieve real-time processing [38]. Processing a 630-frame data set was significantly faster with IS at a fixed IT than SVD filtering (0.83 ± 0.01 s with IS vs 17.3 ± 0.7 s for SVD on a computer equipped with an Intel i7-11700 CPU: 2.5GHz; 8 cores, 16 threads) and 32 GB RAM. However, this advantage could be lost when using automated IT selection, which requires IS filtering to be performed over a range of Its, or if more advanced SVD methods are implemented [39].

Tissue motion compensation was used in this study to increase clutter (tissue) stationarity prior to IS clutter filtering. This improved the efficiency of the interframe subtraction filter and the apparent resolution in both IS and SVD filtered images when motion was present, but had no significant impact on MPD calculated from SVD filtered data (Figure 7 J). This is consistent with qualitative observations that tumors affected by larger motion, visually apparent on cineloops, exhibited larger decreases in MSV following motion compensation. The relatively small and unsignificant increase in MPD post in-plane motion compensation can be attributed to the low sensitivity of MPD to flow velocity (Figure 5 K) and to the automatic adjustment of the rank threshold in SVD processing, which could have contributed to its compensation. Indeed, tissue motion compensation was observed to narrow the singular value peak associated with tissue clutter, while increasing its amplitude. This led to a decrease in the automatic rank from 31.2 ± 20.7 without MC, to 21.9 ± 12.4 after MC, with greater decreases associated with larger visually apparent tumor motion. The implemented tissue motion detection algorithm (‘*imregdemons’*) was selected for its ability to estimate in-plane tissue displacements including potential non-rigid deformations. The extent to which tumors underwent non-rigid deformations was not investigated in this study. It is a time-consuming algorithm (~0.4 s per frame, with the set of parameters used in this study), which preclude real-time processing. It also requires an *a-priori* selected set of parameters, which in this study was empirically determined to compensate tissue motion in the most affected samples while minimizing effect on blood motion in stationary samples, and indistinctively applied to all datasets. Therefore, for real-time tissue motion compensation, other motion estimation algorithms, such as rigid-body registration that has been used with principal component analysis for Doppler ultrasound clutter filtering [25], or block-matching and variants, which are currently widely used in video compression [40], warrant investigation.

The temporal resolution (<10 ms) achieved in this study allowed us to investigate the interframe time on speckle variance in IS clutter filtering using the same set of acquired data. The sensitivity of speckle variance to velocity has two distinct regions (Figure 5 K): an approximately linear increase at lower velocities and a plateau at higher velocities. For blood velocities in the plateau region for a given IT value and acoustic beam correlation length, the blood signal is fully decorrelated and image pairs input to the filter are maximally different, so SV depends only on blood volume. At lower velocities, SV rises with increasing blood velocity and decreasing correlation in the blood signal. Importantly, the range of velocities within this linear variation region decreases with increasing IT, such that the range for the highest IT includes only the sub mm/s velocity range of some capillaries. The vessel phantom and *in vivo* slow flow experiments demonstrate that larger Its are needed to remain sensitive to slower flow signals, for which a greater time is needed for red blood cells to change position and signals to decorrelate. For the smallest IT evaluated here, the range of velocities in the linear region includes blood velocities reported in capillaries, venules and some arterioles [41]. However, the magnitude of the envelope difference is small for slow velocities, yielding a darker visual appearance compared to SVD, with the color scheme used here; other colormaps for image data representation may enhance visualization of pixels with slow flow. The automatically selected IT in the present study should then be interpreted as the interval time above which speckle variance becomes independent of blood velocity in our samples.

The spatial resolution of the proposed flow imaging method was only partly investigated in this work (Supplementary Figure 3). However, when motion compensation performs adequately, the resolution of speckle variance is theoretically similar to that of the B-mode since flow parameters (SV and PD) are estimated along ‘slow time’ (i.e. across registered frames), a dimension which is, in that case, orthogonal to the three spatial dimensions [31,42,43]. It is sufficient to detect flow in the CAM microvasculature and observe its local variations. Super-resolution imaging techniques [44] could eventually be implemented to improve resolution to about capillary diameter and potentially measure blood velocity [45] in these small vessels, but at the cost of a significant increase in both hardware requirements and data acquisition and processing complexity. The manual tumor segmentation for volume measurements constitutes a limitation in the current study, though they were performed by a single user knowledgeable of the anatomy being imaged and experienced in ultrasound imaging. Automated segmentation, when optimized, would certainly be preferable for more extensive studies involving larger numbers of samples and/or multiple data analysts.

To our knowledge, this study also constitutes the first qualitative and quantitative longitudinal investigation of blood flow in a CAM tumor model using UHFUS imaging. The image acquisition method allows a single user to rapidly image dozens of samples within a day and provide longitudinal evaluation of treatment response over a short time-period. In our treatment study, a consistent decrease in tumor MSV was observed in both treatment and control groups in the Renca CAM experiment, a similar outcome to the greater increase in tumor volume compared to vascularity observed following Sorafenib treatment of RCC tumors in mice [46]. MSV was significantly lower after 5 days of treatment (Figure 10 A, B), yet the tumor volume increased and there was no significant difference in fluorescence metrics between groups at study end point, so treatment efficacy is not conclusive in this study. The MPD was also lower at endpoint (Figure 10 B; Supplementary Figure 5), similar to the fractional moving blood volume (FMBV) results in Huang *et al.* [24] where it was also found that tumor growth was not impacted by therapy. Our observation that the MFI was not significantly impacted by therapy is consistent with Huang *et al.* [24], which also reported no difference in microvascular density, as measured with H&E. A notable difference between the present study and Huang *et al.* [24], is that they observed a decrease in fluorescent area (%) with their 7-day, 5 μL/day treatment of 10 μM sunitinib which we did not observe with our 5-day treatment study (10 μL/day of 10 μM sunitinib). We hypothesize that this difference in outcome may be attributable to differences in the treatment regimen and study endpoints. Some studies of RCC tumors in mice have reported increase in survival and decrease in tumor volume and vasculature with longer treatment times [47,48], while other studies have observed treatment resistance to sunitinib that corresponds to vessel co-option. This has been observed in lung metastasis [49,50], which the avian CAM tumor model mimics. The fluorescence histology metrics FA and MFI from the intravenously injected lectin stain are related to perfused microvascular volume whereas the ultrasound metrics MSV and MPD are sensitive to local blood flow. We therefore hypothesize that the differing findings at endpoint reported by the ultrasound metrics (significant) and by the optical metrics (unsignificant) are due to the difference in sensitivity to vessel radius variations. These findings may also indicate that US metrics are sensitive to changes in microvascular flow which could precede and even trigger structural changes observed on histology, such as vessel pruning and normalization [51–53]. However, the validation of this hypothesis is beyond the scope of the present investigation. Future studies can extend the treatment time to further elucidate effects of treatment response and resistance with temporal evaluation since ultrasound imaging inherently enables longitudinal imaging of a specimen. Acoustic shadowing, caused by tissue scarring at the top of the tumor, was evident in some images, making it difficult to assess whether low speckle variance is due to low perfusion, or due to shadowing and depth-dependent attenuation. This can be mitigated by measuring MSV in smaller ROIs, avoiding areas with low speckle brightness in B-mode, rather than in a ROI encompassing the entire tumor.

IS filtering with tissue motion compensation can be easily implemented with offline data processing by acquiring uncompressed image data using the RF-mode, or decompressing RAW data acquired in standard mode. This approach can enable either automatic IT selection or for variation of MSV with IT to be analyzed. Online IS filtering and real-time display of speckle variance images requires the data processing and image display pipeline on the imaging platform to be modified in an engineering or programming mode on the scanner. In this case, a single IT would be selected on the fly by the user, in the same way clutter filtering parameters and velocity range are selected in color Doppler imaging. The frame ensemble size would also be significantly reduced compared to that used in the present study (i.e. $N_{max} = 630$), at no cost with regard to IS clutter filtering, but with a reduction in speckle variance precision, which could potentially affect microvessels detection. This would warrant further investigations.

Here we presented a method for rapid and non-invasive microvasculature imaging and blood flow quantification in the CAM tumor model. We demonstrated that IS with tissue motion compensation is as capable as SVD of filtering out clutter signal for flow detection in the microvasculature. Vessel phantom experiments demonstrated that by varying IT one can target various flow velocities, with slower velocities being detected at larger IT. In CAM experiments, non-rigid motion compensation coupled with frame suppression significantly reduced MSV which suggests that tissue motion indeed affects blood flow quantification and must be compensated for prior to IS filtering. Lastly, due to its computational simplicity and efficiency, IS is an attractive clutter filtering method to be implemented in a high throughput clinical tool such as the CAM PDX tumor model for evaluating drug regimens.

By using widely available equipment and computational power levels, we have developed a non-invasive, accessible, and cost-effective clutter filter based UHFUS imaging method which allows for the quantification of microvascular flow in avian-grafted tumors, fulfilling an unmet need for providing vascularity metrics. The findings demonstrated here may have significant clinical implications as they suggest the potential for this method for assessing the efficacy of anti-angiogenic agents in xenografts from patients with highly angiogenic cancers, which would in turn have potential to increase objective response rates and improve patient survival.

**Acknowledgments**:
We would like to thank Olivia Grafinger for her contribution to the introduction.
Funding for these investigations was provided by the Canadian Institute of Health Research (Foundation Grant 148367 and Project Grant 156131), and Justin Xu received an Undergraduate Student Research Award from the University of Toronto.

**References:**

1. DeBord LC, Pathak RR, Villaneuva M, Liu H-C, Harrington DA, Yu W, et al. The chick chorioallantoic membrane (CAM) as a versatile patient-derived xenograft (PDX) platform for precision medicine and preclinical research. Am J Cancer Res. 2018;8: 1642–1660.

2. Hu J, Ishihara M, Chin AI, Wu L. Establishment of xenografts of urological cancers on chicken chorioallantoic membrane (CAM) to study metastasis. Precis Clin Med. 2019;2: 140–151.

3. Tran TA, Leong HS, Pavia-Jimenez A, Fedyshyn S, Yang J, Kucejova B, et al. Fibroblast Growth Factor Receptor-Dependent and -Independent Paracrine Signaling by Sunitinib-Resistant Renal Cell Carcinoma. Mol Cell Biol. 2016;36: 1836–1855.

4. Charbonneau M, Harper K, Brochu-Gaudreau K, Perreault A, McDonald PP, Ekindi-Ndongo N, et al. Establishment of a ccRCC patient-derived chick chorioallantoic membrane model for drug testing. Front Med. 2022;9: 1003914.

5. Hao Z, Sadek I. Sunitinib: the antiangiogenic effects and beyond. Onco Targets Ther. 2016;9: 5495–5505.

6. Comandone A, Vana F, Comandone T, Tucci M. Antiangiogenic therapy in clear cell renal carcinoma (CCRC): Pharmacological basis and clinical results. Cancers (Basel). 2021;13: 5896.

7. Catalano M, Procopio G, Sepe P, Santoni M, Sessa F, Villari D, et al. Tyrosine kinase and immune checkpoints inhibitors in favorable risk metastatic renal cell carcinoma: Trick or treat? Pharmacol Ther. 2023;249: 108499.

8. Deleuze A, Saout J, Dugay F, Peyronnet B, Mathieu R, Verhoest G, et al. Immunotherapy in Renal Cell Carcinoma: The Future Is Now. Int J Mol Sci. 2020;21. doi:10.3390/ijms21072532

9. Canil C, Kapoor A, Basappa NS, Bjarnason G, Bossé D, Dudani S, et al. Management of advanced kidney cancer: Kidney Cancer Research Network of Canada (KCRNC) consensus update 2021. Can Urol Assoc J. 2021;15: 84–97.

10. Khan KA, Kerbel RS. Improving immunotherapy outcomes with anti-angiogenic treatments and vice versa. Nat Rev Clin Oncol. 2018;15: 310–324.

11. Motzer RJ, Rini BI, McDermott DF, Redman BG, Kuzel TM, Harrison MR, et al. Nivolumab for Metastatic Renal Cell Carcinoma: Results of a Randomized Phase II Trial. J Clin Oncol. 2015;33: 1430–1437.

12. Motzer RJ, Escudier B, McDermott DF, George S, Hammers HJ, Srinivas S, et al. Nivolumab versus Everolimus in Advanced Renal-Cell Carcinoma. N Engl J Med. 2015;373: 1803–1813.

13. Yip SM, Wells C, Moreira R, Wong A, Srinivas S, Beuselinck B, et al. Checkpoint inhibitors in patients with metastatic renal cell carcinoma: Results from the International Metastatic Renal Cell Carcinoma Database Consortium. Cancer. 2018;124: 3677–3683.

14. McDermott DF, Drake CG, Sznol M, Choueiri TK, Powderly JD, Smith DC, et al. Survival, Durable Response, and Long-Term Safety in Patients With Previously Treated Advanced Renal Cell Carcinoma Receiving Nivolumab. J Clin Oncol. 2015;33: 2013–2020.

15. Goertz D, Yu J, Kerbel R, Burns P, Foster F. High-frequency Doppler ultrasound monitors the effects of antivascular therapy on tumor blood flow. Cancer Res. 2002;62: 6371–6375.

16. Graham KC, Wirtzfeld LA, MacKenzie LT, Postenka CO, Groom AC, MacDonald IC, et al. Three-dimensional high-frequency ultrasound imaging for longitudinal evaluation of liver metastases in preclinical models. Cancer Res. 2005;65: 5231–5237.

17. Vitiello M, Kusmic C, Faita F, Poliseno L. Analysis of Lymph Node Volume by Ultra-High-Frequency Ultrasound Imaging in the Braf/Pten Genetically Engineered Mouse Model of Melanoma. J Vis Exp. 2021. doi:10.3791/62527

18. Martin KH, Lindsey BD, Ma J, Nichols TC, Jiang X, Dayton PA. Ex Vivo Porcine Arterial and Chorioallantoic Membrane Acoustic Angiography Using Dual-Frequency Intravascular Ultrasound Probes. Ultrasound in Medicine and Biology. 2015;42: 2294–2307.

19. Lowerison MR, Willie CJ, Pardhan S, Power NE, Chambers AF, Leong HS, et al. Abstract B02: Ultrasound evaluation of anti-angiogenic therapy on patient-derived renal cell carcinoma xenograft tumors in the chicken embryo model. Mol Cancer Ther. American Association for Cancer Research; 2015. pp. B02–B02.

20. Lowerison MR. First-order statistical speckle models improve robustness and reproducibility of contrast-enhanced ultrasound perfusion estimates. PhD Thesis, The University of Western Ontario. 2017. Available: https://ir.lib.uwo.ca/etd/4399

21. Eckrich J, Kugler P, Buhr CR, Ernst BP, Mendler S, Baumgart J, et al. Monitoring of tumor growth and vascularization with repetitive ultrasonography in the chicken chorioallantoic-membrane-assay. Sci Rep. 2020;10: 18585.

22. Lok U-W, Trzasko JD, Huang C, Tang S, Gong P, Kim Y, et al. Improved Ultrasound Microvessel Imaging Using Deconvolution with Total Variation Regularization. Ultrasound Med Biol. 2021;47: 1089–1098.

23. Pinter SZ, Lacefield JC. Detectability of small blood vessels with high-frequency power Doppler and selection of wall filter cut-off velocity for microvascular imaging. Ultrasound Med Biol. 2009;35: 1217–1228.

24. Huang C, Lowerison MR, Lucien F, Gong P, Wang D, Song P, et al. Noninvasive Contrast-Free 3D Evaluation of Tumor Angiogenesis with Ultrasensitive Ultrasound Microvessel Imaging. Sci Rep. 2019;9: 4907.

25. Babaei S, Dai B, Abbey CK, Ambreen Y, Dobrucki WL, Insana MF. Monitoring muscle perfusion in rodents during short-term ischemia using power Doppler ultrasound. Ultrasound Med Biol. 2023;49: 1465–1475.

26. Baranger J, Arnal B, Perren F, Baud O, Tanter M, Demene C. Adaptive Spatiotemporal SVD Clutter Filtering for Ultrafast Doppler Imaging Using Similarity of Spatial Singular Vectors. IEEE Trans Med Imaging. 2018;37: 1574–1586.

27. Riemer K, Lerendegui M, Toulemonde M, Zhu J, Dunsby C, Weinberg PD, et al. On the use of Singular Value Decomposition as a clutter filter for ultrasound flow imaging. arXiv [physics.med-ph]. 2023. doi:10.48550/arXiv.2304.12783

28. Rausch M, Weiss A, Zoetemelk M, Piersma SR, Jimenez CR, van Beijnum JR, et al. Optimized Combination of HDACI and TKI Efficiently Inhibits Metabolic Activity in Renal Cell Carcinoma and Overcomes Sunitinib Resistance. Cancers . 2020;12. doi:10.3390/cancers12113172

29. Digital RF-Mode. In: FUJIFILM Visualsonics [Internet]. [cited 9 Jan 2026]. Available: https://www.visualsonics.com/product/software/digital-rf-mode

30. Needles A, Goertz DE, Cheung AM, Foster FS. Interframe clutter filtering for high frequency flow imaging. Ultrasound Med Biol. 2007;33: 591–600.

31. Rubin JM, Tuthill TA, Fowlkes JB. Volume flow measurement using Doppler and grey-scale decorrelation. Ultrasound Med Biol. 2001;27: 101–109.

32. Demene C, Deffieux T, Pernot M, Osmanski B-F, Biran V, Gennisson J-L, et al. Spatiotemporal Clutter Filtering of Ultrafast Ultrasound Data Highly Increases Doppler and fUltrasound Sensitivity. IEEE Trans Med Imaging. 2015;34: 2271–2285.

33. Andrewartha SJ, Tazawa H, Burggren WW. Embryonic control of heart rate: examining developmental patterns and temperature and oxygenation influences using embryonic avian models. Respir Physiol Neurobiol. 2011;178: 84–96.

34. Mortola JP, Wills K, Trippenbach T, Al Awam K. Interactive effects of temperature and hypoxia on heart rate and oxygen consumption of the 3-day old chicken embryo. Comp Biochem Physiol A Mol Integr Physiol. 2010;155: 301–308.

35. Jilani SM, Murphy TJ, Thai SNM, Eichmann A, Alva JA, Iruela-Arispe ML. Selective binding of lectins to embryonic chicken vasculature. J Histochem Cytochem. 2003;51: 597–604.

36. Rasche C. Rapid contour detection for image classification. IET Image Process. 2018;12: 532–538.

37. Adler RS, Rubin JM, Fowlkes JB, Carson PL, Pallister JE. Ultrasonic estimation of tissue perfusion: a stochastic approach. Ultrasound Med Biol. 1995;21: 493–500.

38. Lahabar S, Narayanan PJ. Singular value decomposition on GPU using CUDA. 2009 IEEE International Symposium on Parallel & Distributed Processing. IEEE; 2009. pp. 1–10.

39. Lok U-W, Song P, Trzasko JD, Daigle R, Borisch EA, Huang C, et al. Real time SVD-based clutter filtering using randomized singular value decomposition and spatial downsampling for micro-vessel imaging on a Verasonics ultrasound system. Ultrasonics. 2020;107: 106163.

40. Lixin Z, Xuecheng Z, Weizhong L. A fast block-matching motion estimation algorithm for H.264/AVC. 2006 6th International Conference on ITS Telecommunications. IEEE; 2006. pp. 1289–1292.

41. Alves de Mesquita J Jr, Bouskela E, Wajnberg E, Lopes de Melo P. Improved instrumentation for blood flow velocity measurements in the microcirculation of small animals. Rev Sci Instrum. 2007;78: 024303.

42. Vray D, Needles A, Yang VXD, Foster FS. High frequency b-mode ultrasound blood flow estimation in the microvasculature. IEEE Ultrasonics Symposium, 2004. IEEE; 2005. doi:10.1109/ultsym.2004.1417763

43. Yang VXD, Needles A, Vray D, Lo S, Wilson BC, Vitkin IA, et al. High frequency ultrasound speckle flow imaging - comparision with doppler optical coherence tomography (DOCT). IEEE Ultrasonics Symposium, 2004. IEEE; 2005. doi:10.1109/ultsym.2004.1417760

44. Jensen JA, Naji MA, Praesius SK, Taghavi I, Schou M, Hansen LN, et al. Super Resolution Ultrasound Imaging using the Erythrocytes: I: Density Images. IEEE Trans Ultrason Ferroelectr Freq Control. 2024;PP. doi:10.1109/TUFFC.2024.3411711

45. Naji MA, Taghavi I, Schou M, Praesius SK, Hansen LN, Panduro NS, et al. Super Resolution Ultrasound Imaging Using the Erythrocytes: II: Velocity Images. IEEE Trans Ultrason Ferroelectr Freq Control. 2024;PP. doi:10.1109/TUFFC.2024.3411795

46. Schor-Bardach R, Alsop DC, Pedrosa I, Solazzo SA, Wang X, Marquis RP, et al. Does arterial spin-labeling MR imaging-measured tumor perfusion correlate with renal cell cancer response to antiangiogenic therapy in a mouse model? Radiology. 2009;251: 731–742.

47. Welti JC, Powles T, Foo S, Gourlaouen M, Preece N, Foster J, et al. Contrasting effects of sunitinib within in vivo models of metastasis. Angiogenesis. 2012;15: 623–641.

48. Hara T, Miyake H, Hinata N, Fujisawa M. Inhibition of tumor growth and sensitization to sunitinib by RNA interference targeting programmed death-ligand 1 in mouse renal cell carcinoma RenCa model. Anticancer Res. 2019;39: 4737–4742.

49. Bridgeman VL, Vermeulen PB, Foo S, Bilecz A, Daley F, Kostaras E, et al. Vessel co-option is common in human lung metastases and mediates resistance to anti-angiogenic therapy in preclinical lung metastasis models: Vessel co-option in lung metastases. J Pathol. 2017;241: 362–374.

50. Lichner Z, Saleeb R, Butz H, Ding Q, Nofech-Mozes R, Riad S, et al. Sunitinib induces early histomolecular changes in a subset of renal cancer cells that contribute to resistance. FASEB J. 2019;33: 1347–1359.

51. Pries AR, Cornelissen AJM, Sloot AA, Hinkeldey M, Dreher MR, Höpfner M, et al. Structural adaptation and heterogeneity of normal and tumor microvascular networks. PLoS Comput Biol. 2009;5: e1000394.

52. Pries AR, Secomb TW. Making microvascular networks work: angiogenesis, remodeling, and pruning. Physiology (Bethesda). 2014;29: 446–455.

53. Korn C, Augustin HG. Mechanisms of vessel pruning and regression. Dev Cell. 2015;34: 5–17.

**Abbreviations:**
CAM: chorioallantoic membrane
E: ultrasound signal envelope
ED: embryonic day
FA: relative fluorescent area
FR: frame rate
IQ: I (in-phase) and Q (quadrature) components of ultrasound signals
IS: interframe subtraction
IT: interframe time
MC: motion compensation
MFI: mean fluorescence intensity in a region of interest
mRCC: metastatic renal cell carcinoma
M: mean
MPD: mean power Doppler in a region of interest

MSV: mean speckle temporal variance in a region of interest
PD: power Doppler
PDX: patient derived xenograft
RAW data: logarithmically compressed envelope data
RCC: renal cell carcinoma
ROI: region of interest
SD: standard deviation
SV: speckle temporal variance
SVD: singular value decomposition
UHFUS: ultra-high frequency ultrasound
UMI: ultrasound microvessel imaging

**Conflicts of interest:**
The authors declare that they have no competing interests.

**Figures:**

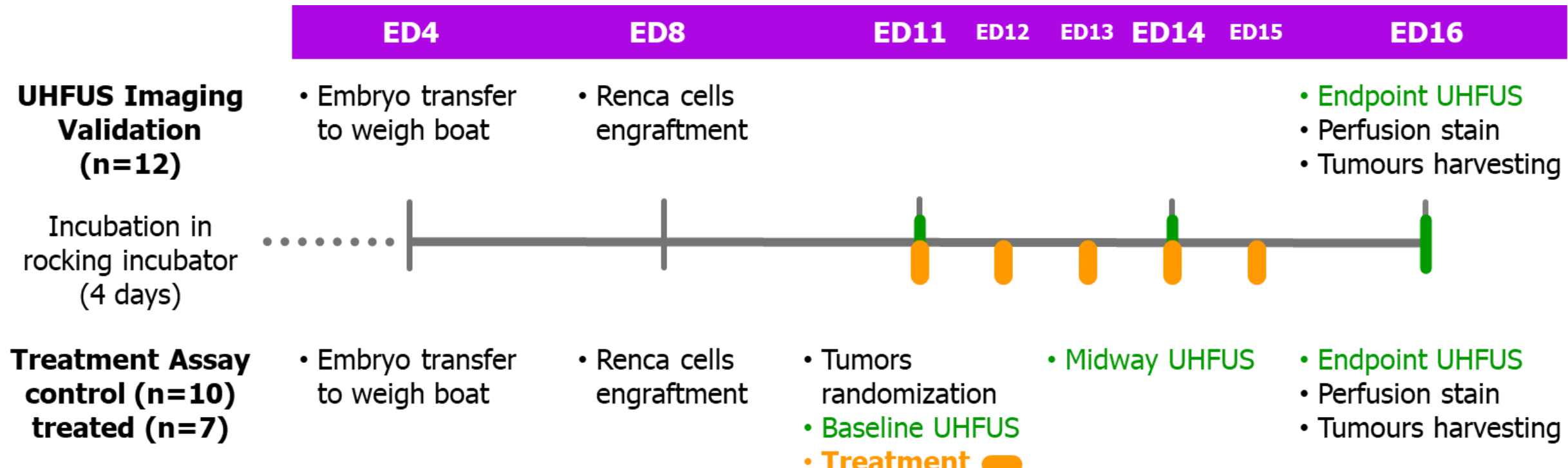

Figure 1: Experimental timelines for CAM tumor model imaging experiment (top) and treatment assay (bottom).

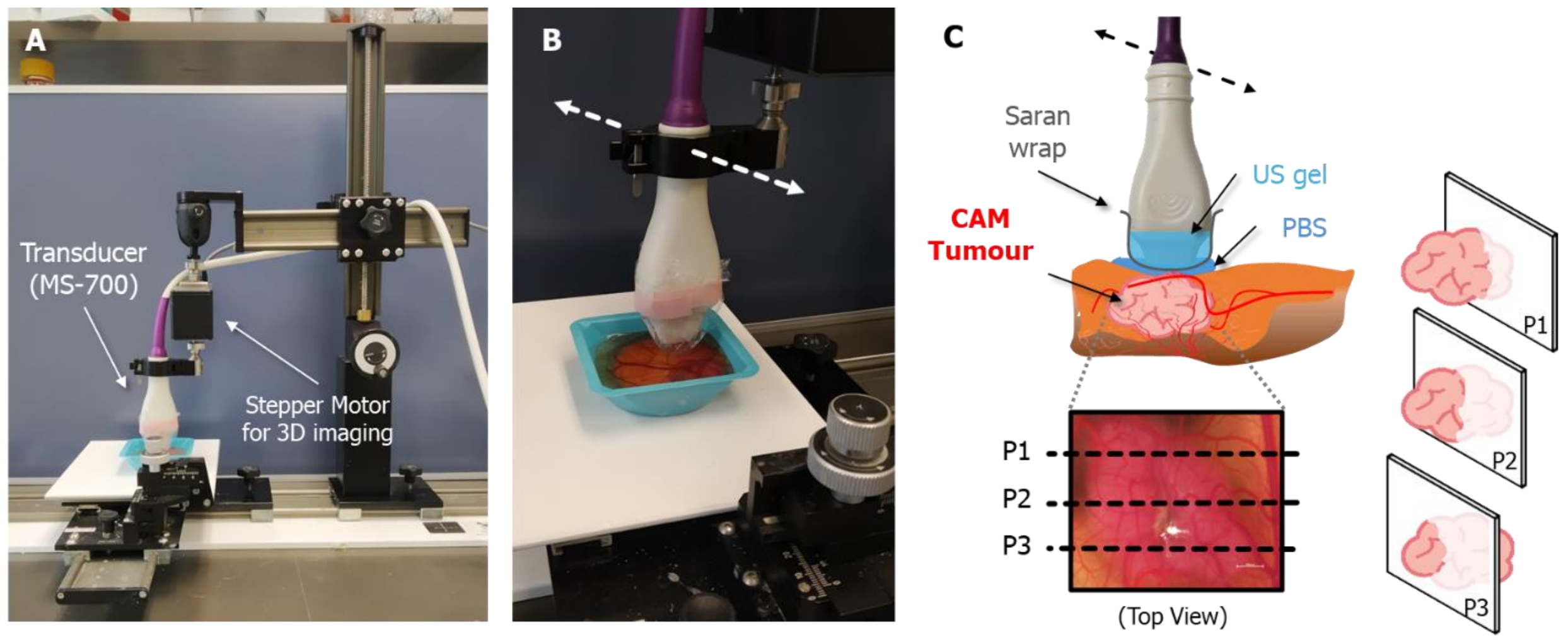

Figure 2: MS-700 transducer set up over embryo in weigh boat (A, B). Schematic of UHFUS imaging protocol at three planes (C).

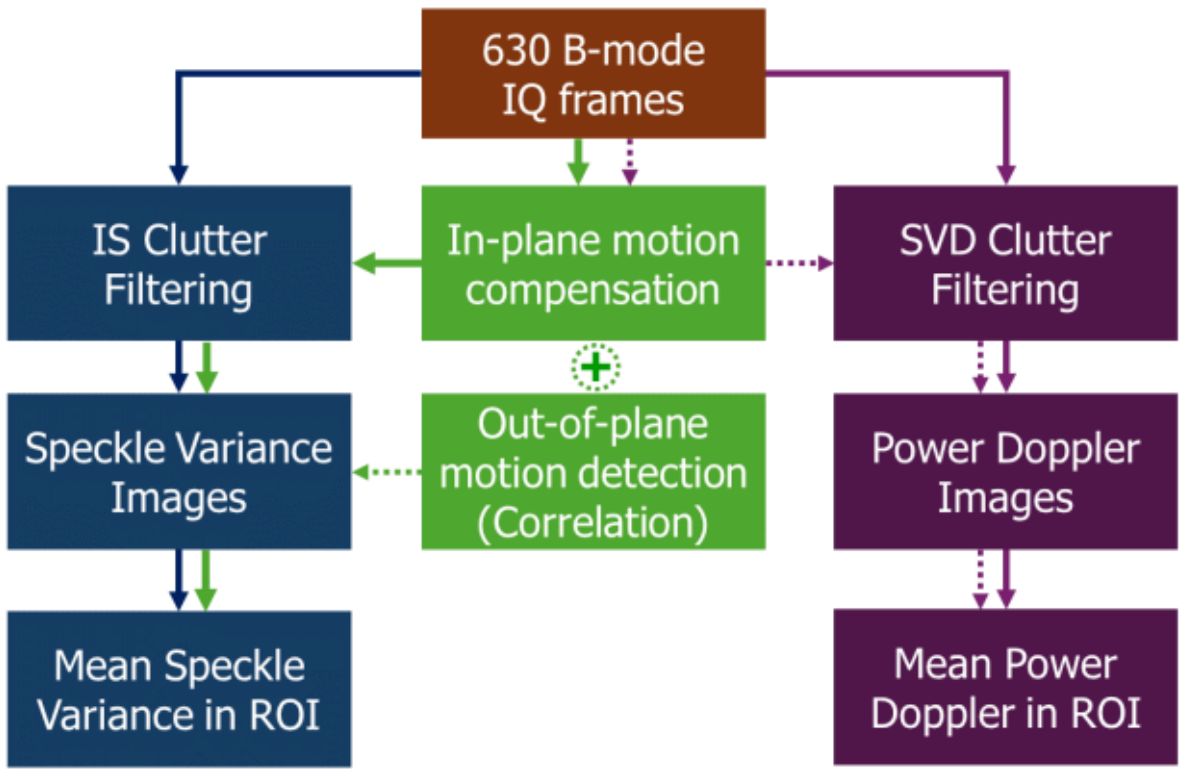

Figure 3: UHFUS data acquisition and processing pipeline: IS filtering with (green path) and without (blue path) tissue motion compensation. UMI (purple) with (dashed) and without (plain) tissue motion compensation.

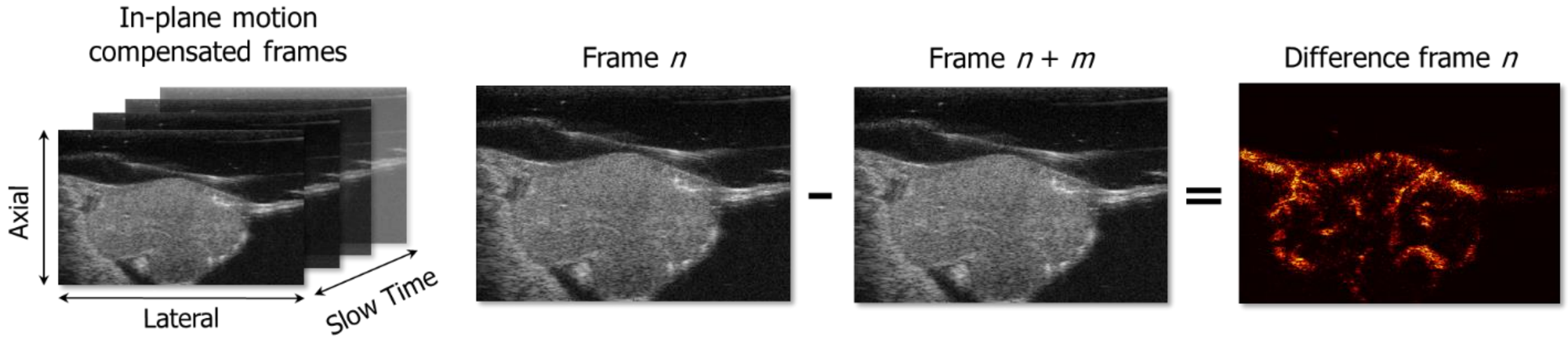

Figure 4: Interframe subtraction clutter filter based on subtraction of frames acquired at the same location at a given interframe time (IT) interval.

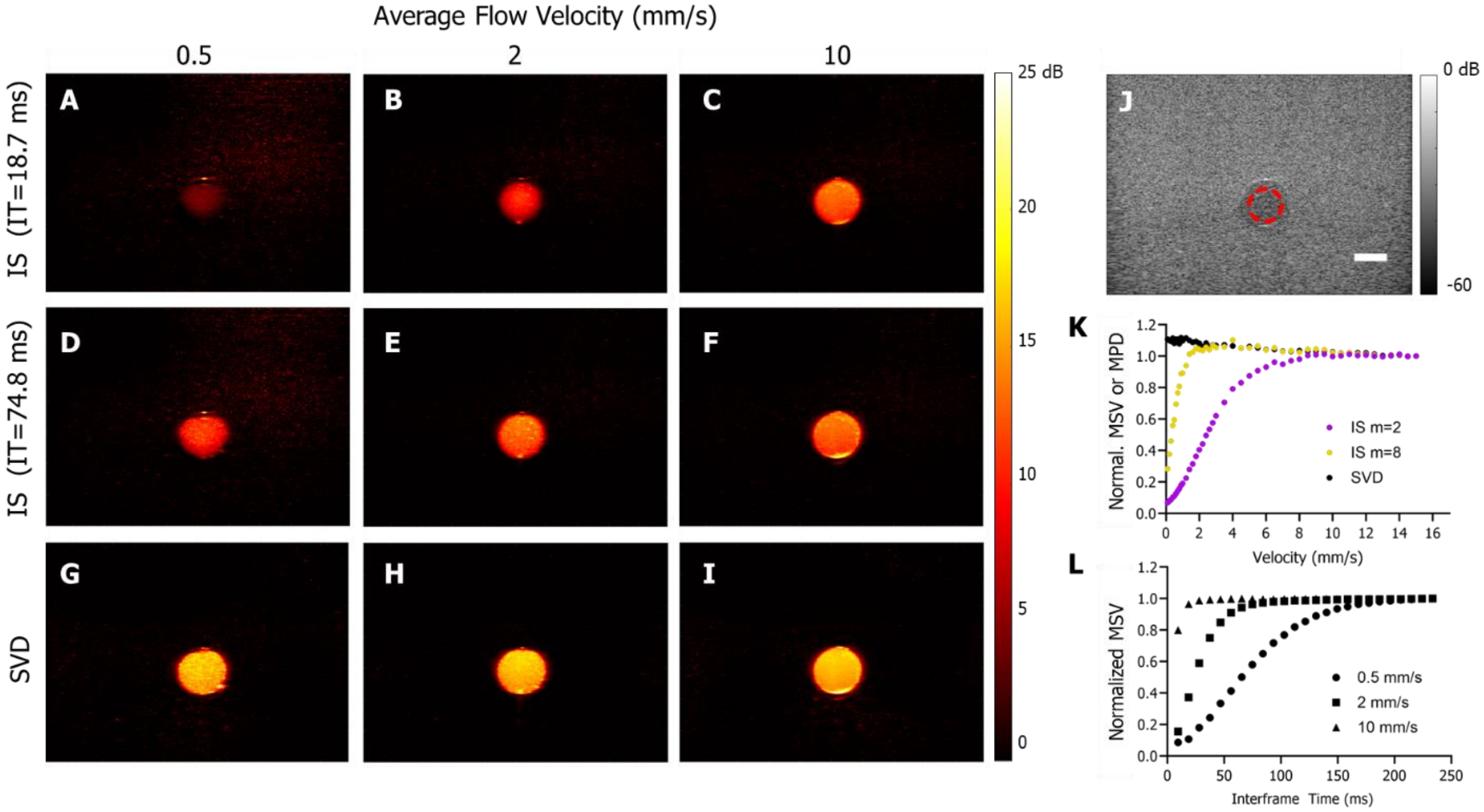

Figure 5: Log-compressed speckle variance or power Doppler images of the cross section of the vessel phantom for mean flow velocity of 0.5 mm/s, 2 mm/s and 10 mm/s, processed without MC (dynamic range =25dB) and either IS filtering with *m*=2 (IT=18.7 ms; A-C), *m*=8 (IT=74.8 ms; D-F) or SVD filtering (G-I). B-mode image of a cross section of the vessel phantom, with ROI for MSV and MPD calculation indicated by dashed circle; scale bar: 1 mm (J). Mean speckle variance as a function of mean flow velocity, after IS filtering with IT=18.7 ms (*m*=2) and IT=74.8 ms (*m*=8), and mean power Doppler after SVD filtering (both normalized to their value at the highest flow velocity) (K). Normalized MSV as a function of IT for mean flow velocity of 0.5, 2 and 10 mm/s (L).

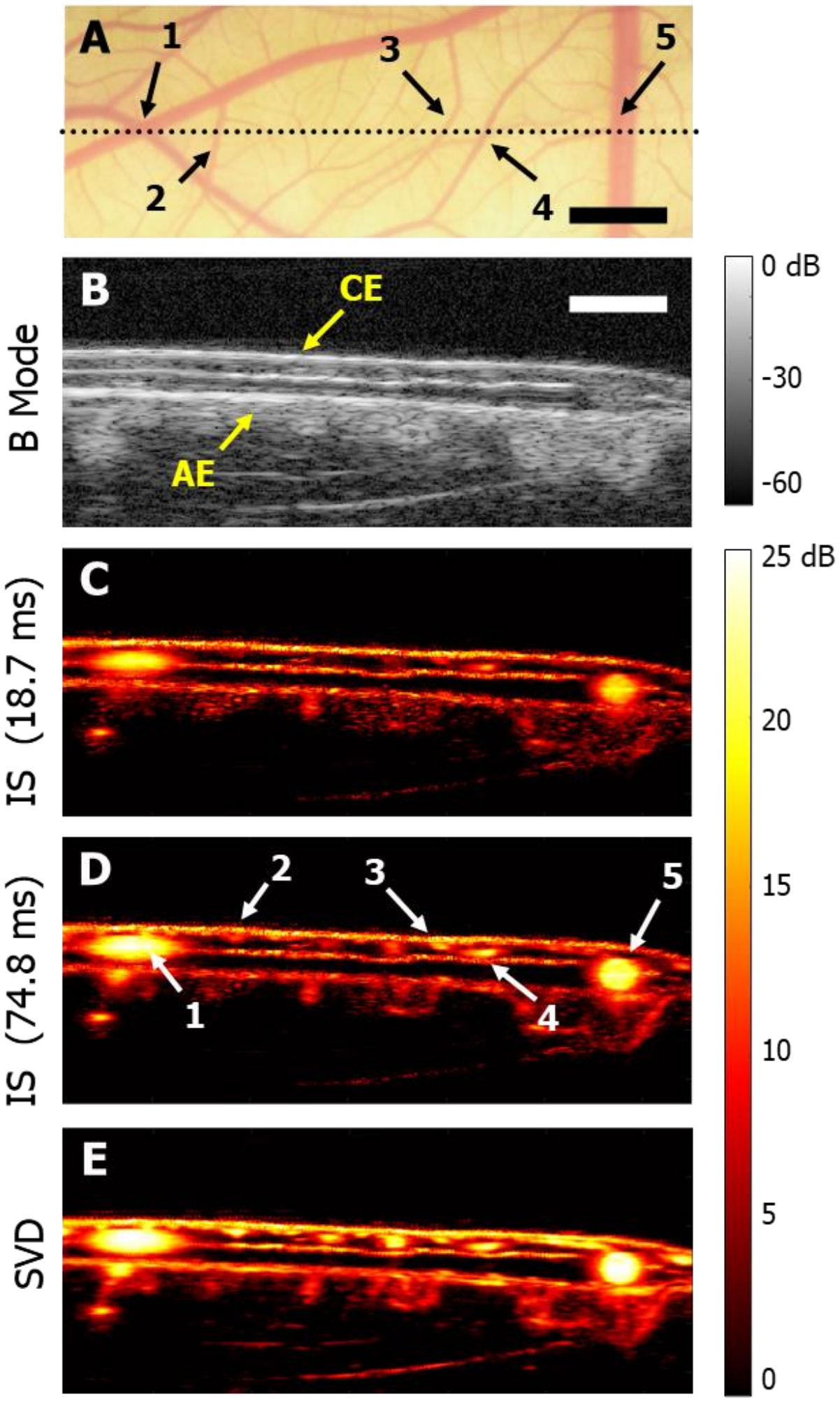

Figure 6: Brightfield image of a CAM vessel bed (A), with the dashed line representing the μUS imaging plane. B-mode image of the CAM vessel bed with the chorionic (CE) and allantoic (AE) epithelia (B). Log-compressed speckle variance images obtained post tissue motion compensation and IS clutter filtering (*m*=2 / IT=18.7 ms; C) and IS (*m*=8 / IT=74.8 ms; D). Power Doppler image, post tissue motion compensation and SVD clutter filtering (E). Vessel diameters of 178 μm (1), 80 μm (2), 60 μm (3), 72 μm (4) and 251 μm (5) were measured on the brightfield image. Scale bars: 1 mm.

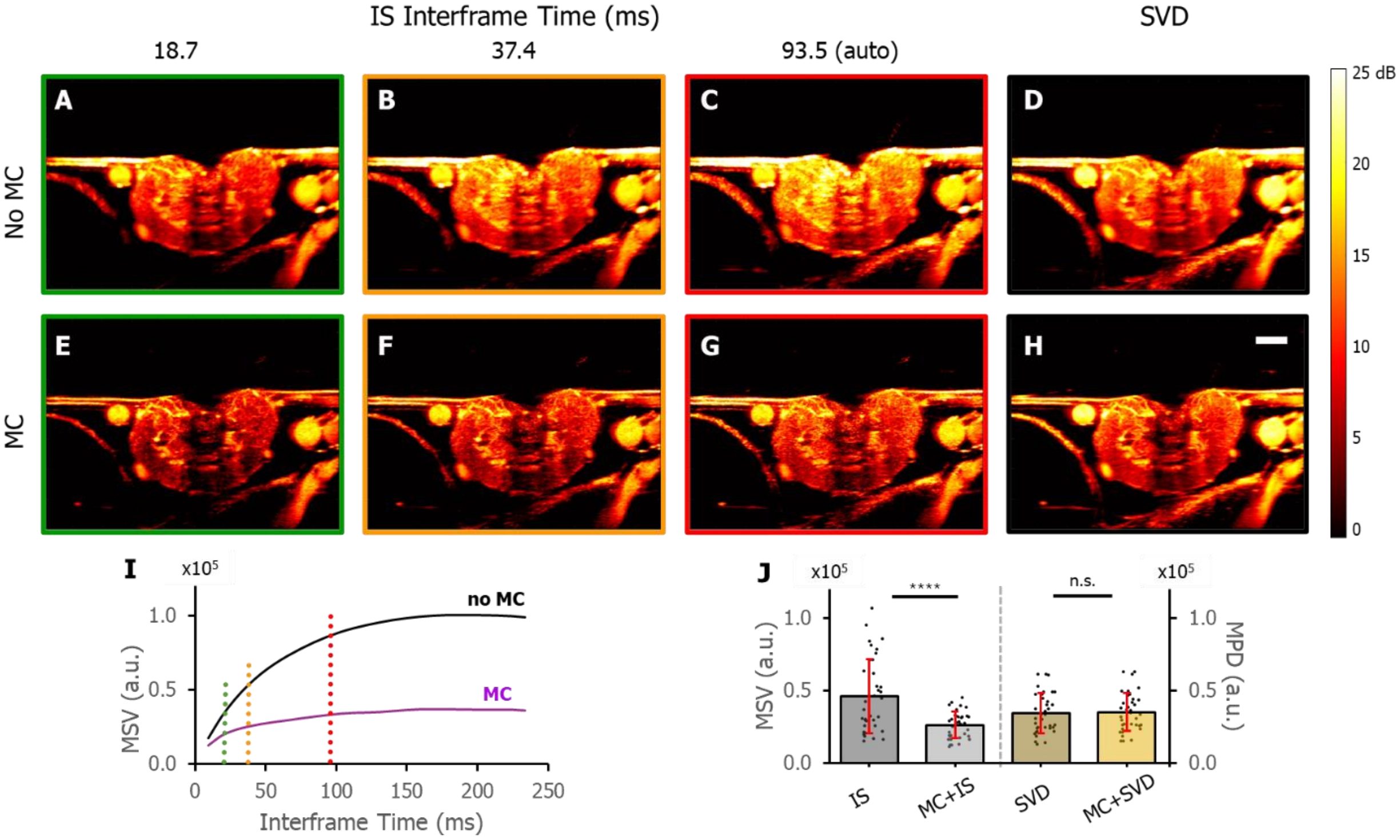

Figure 7: Log-compressed speckle variance images of a CAM Renca tumor obtained without (A-C) and with tissue motion compensation (E-G), and IS filtering with IT=18.7 ms (m=2; A, E), 37.4 ms (m=4; B, F), 93.5 ms (m=10, auto-selected; C, G); Log-compressed power Doppler image obtained with (D) and without tissue motion compensation (H; Scale bar: 1 mm.) and SVD filtering. Mean speckle variance measured in the tumor ROI (dashed line in A) as a function of IS interframe time, with and without prior motion compensation (I). Evaluation of the effect of motion compensation on the mean speckle variance (post IS filtering with auto IT) and mean power Doppler (post SVD filtering) measured in 3 planes of 12 Renca tumors (****: $p<10^{-5}$, n.s.: p>0.1) Error bars denote standard deviation) (J).

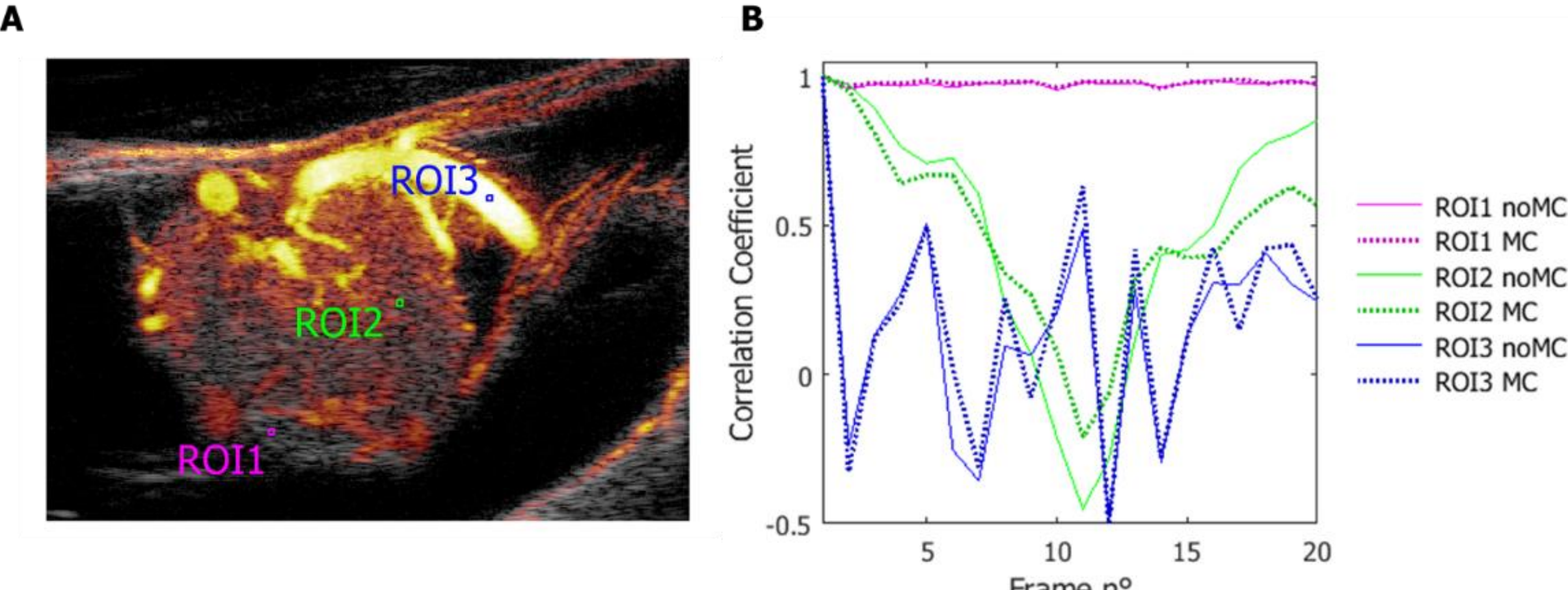

Figure 8: Effect of motion compensation (MC) on moving blood signal in an otherwise visually stationary tumor. (A) Speckle variance image overlaid on top of the first B-mode cineloop frame with three 60μm x 54μm ROI selected in areas of various flow conditions (ROI1: no apparent flow, ROI2: 'medium' flow, ROI3: 'fast' flow). (B) Comparison of the correlation coefficients between the envelope signal measured in these ROIs in frames 2 to 20 and that in frame 1 obtained, without (solid line) and with (dashed line) MC.

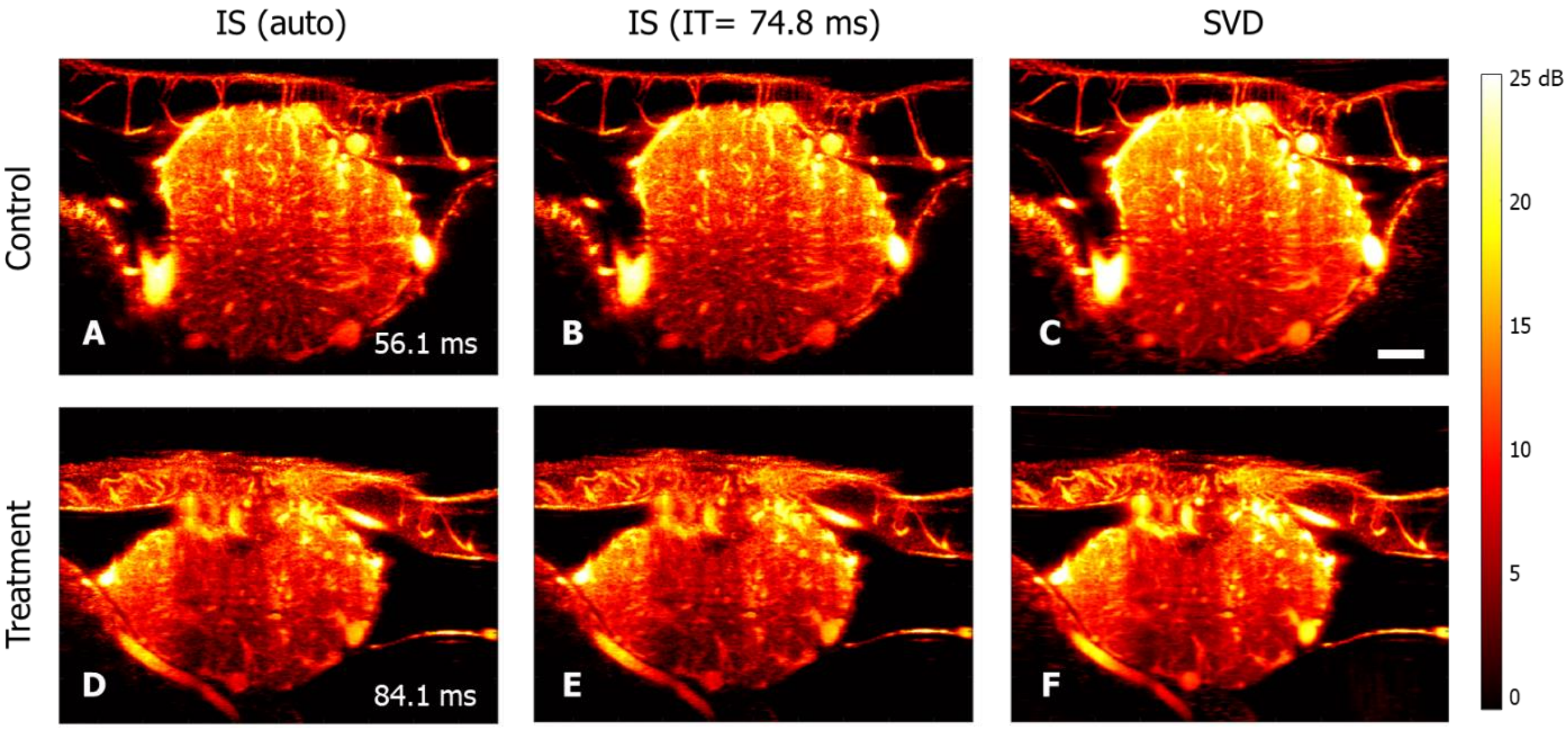

Figure 9: Speckle variance (A-B, D-E) and power Doppler (C, F) images of control and treated Renca CAM tumors obtained at ED 16. Tissue motion compensation was used prior to IS (A-B, D-E) and SVD (C, F) clutter filtering. For IS filtering, the auto-selected IT is reported in the lower right corner. Scale bar: 1 mm.

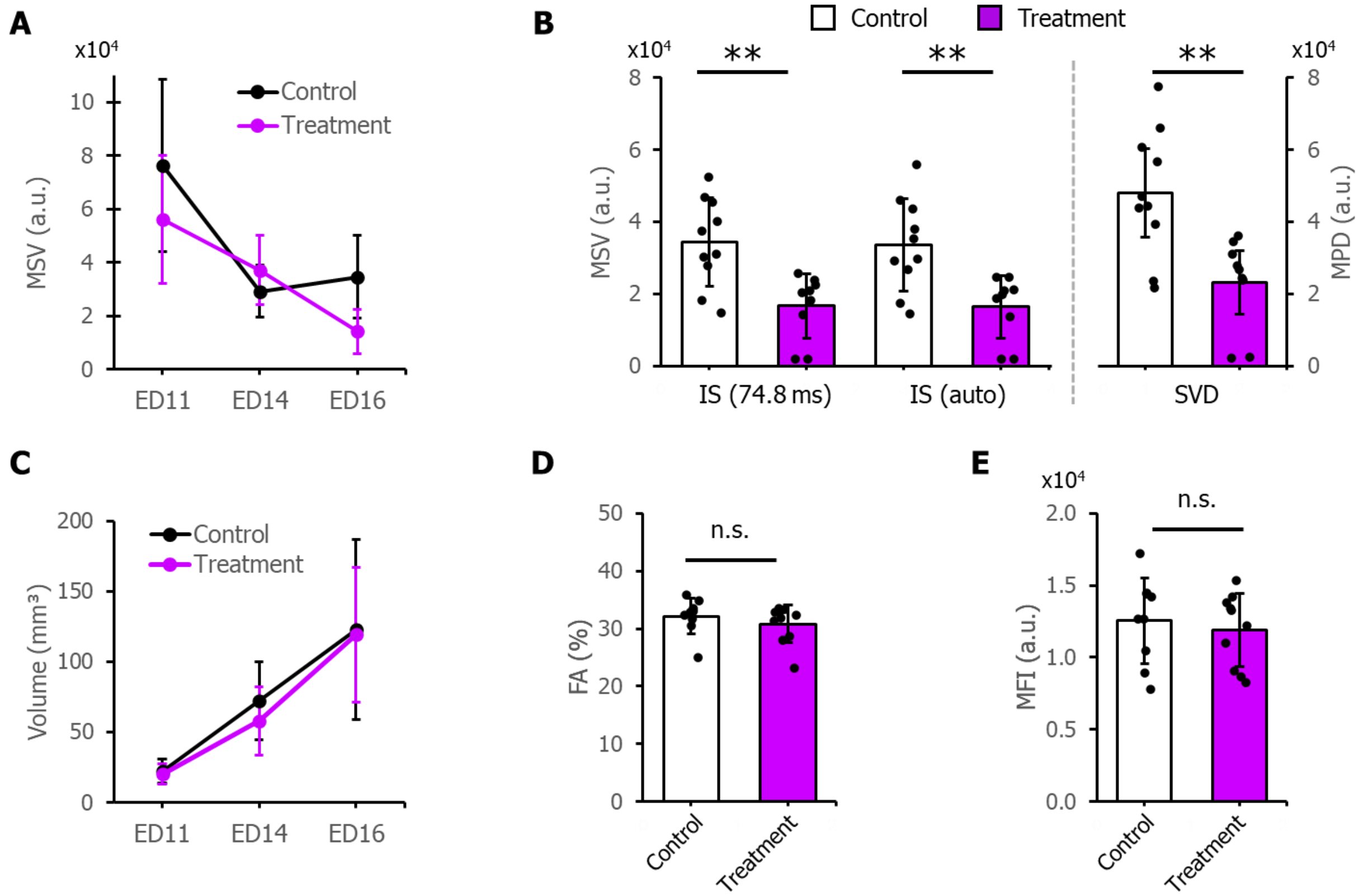

Figure 10: Mean speckle variance measured throughout experiment timeline in control (n=10) and treated (n=11, 10 and 9 at ED 11, ED 14 and ED 16, respectively) tumors post tissue motion compensation and IS filtering (IT=74.8 ms) (A). Mean speckle variance measured post MC and IS filtering (IT=74.8 ms and 'auto') and mean power Doppler measured in the same groups at endpoint (**: $p<10^{-2}$, n.s.: $p>0.3$) (B). Mean tumor volume estimated from 3-D UHFUS acquisitions at baseline, midway and endpoint (C). Fluorescent area percentage in perfusion stain sections measured at endpoint (D). Mean Fluorescence Intensity in perfusion stain sections at end point (E). Error bars denote standard deviation.

**Supplementary Figures** :

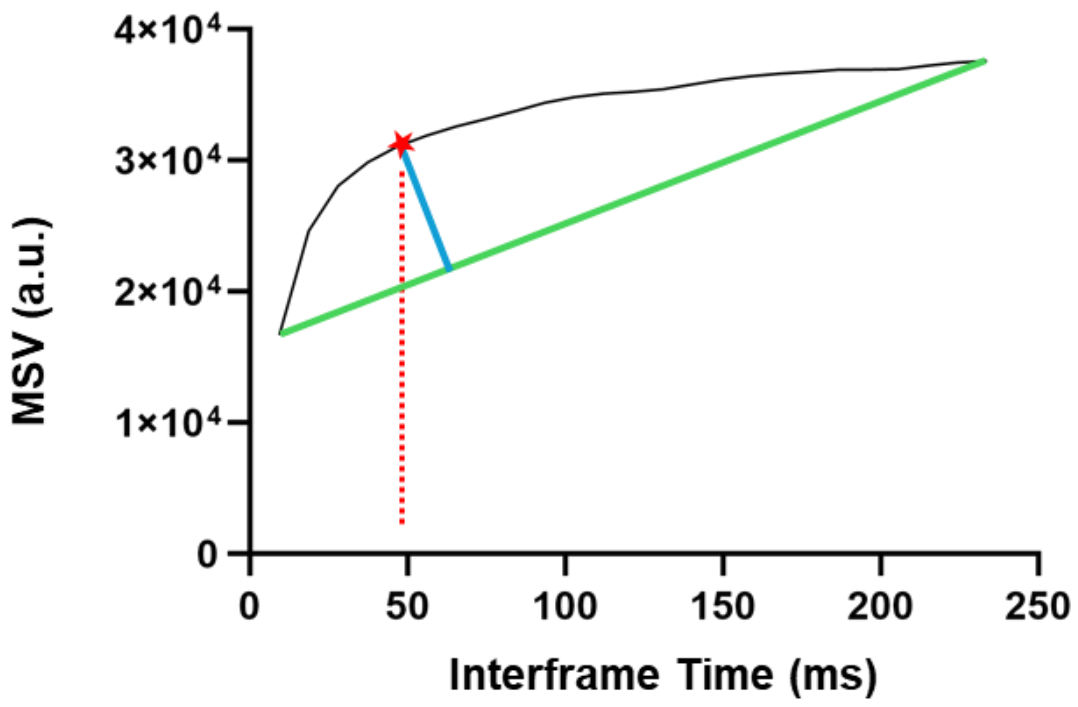

Supplementary Figure 1: Automatic interframe time selection based on the maximum perpendicular distance (blue) between mean speckle variance (black) and a line joining the first and last MSV values (green). For this tumor, an IT of 46.73 ms (i.e., m=5) was selected.

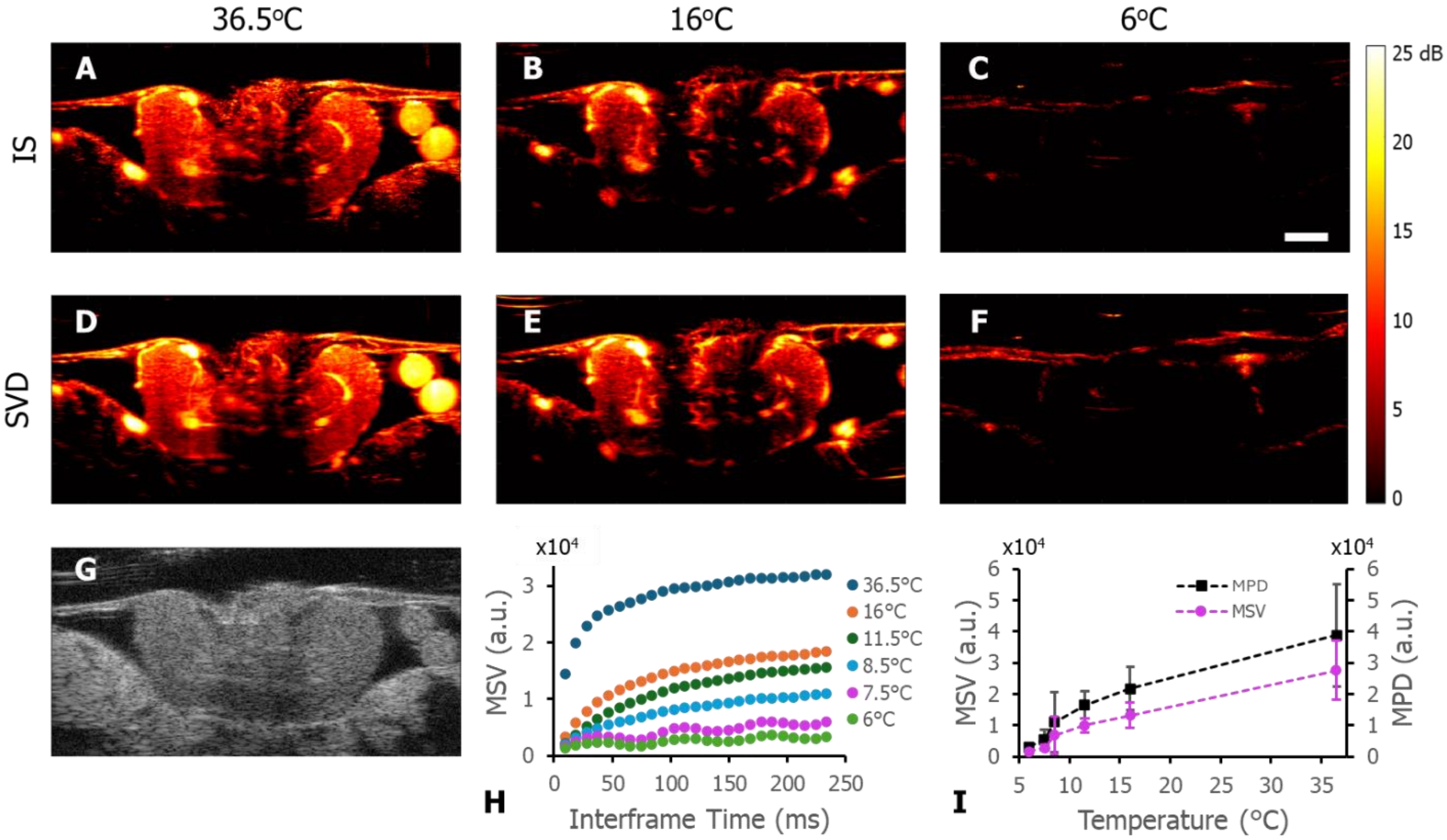

Supplementary Figure 2: Log-compressed speckle variance images of a CAM Renca tumor obtained at various embryo temperatures: 36.5°C (A), 16°C (B) and 6°C (C), post IS filtering (*m*=8 / IT=74.8 ms; scale bar: 1 mm). Log-compressed power Doppler images, post SVD filtering at the same embryo temperatures (D-F). B-mode image of the tumor (G). Mean speckle variance (MC+IS) as a function of IT at each embryo body temperature (H). Mean speckle variance (post IS, IT=74.8 ms) and mean power Doppler (post SVD) as a function of body temperature (I). Data in all panels were obtained post tissue motion compensation.

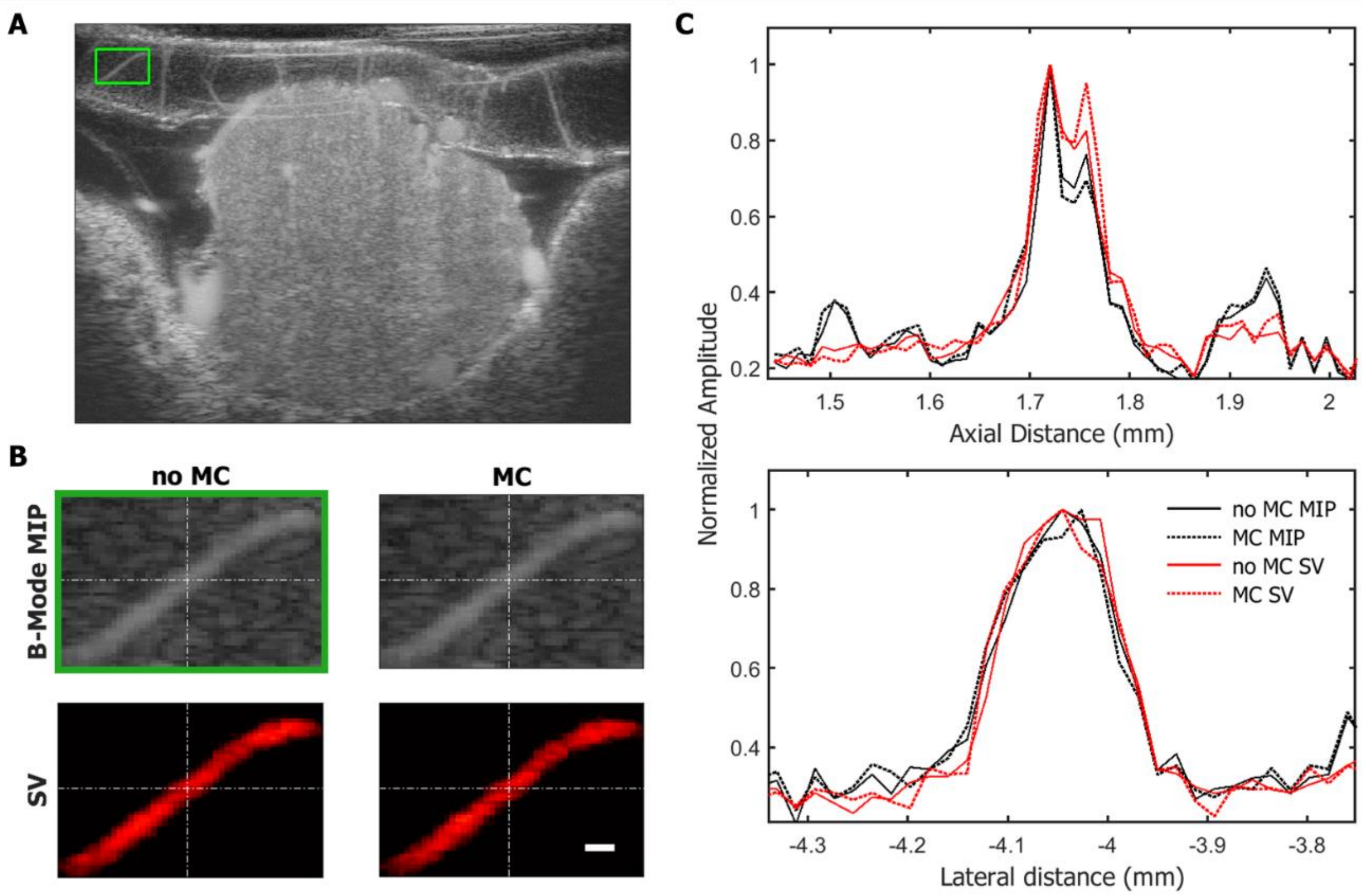

Supplementary Figure 3: Comparison of B-mode and SV image resolutions; (A) Maximum intensity projection (MIP) image from 630 B-mode frames (not motion-corrected) of a 'stationary' tumor showing microvessels in the CAM above the tumor; (B) images of MIP and SV calculated with and without motion correction in the ROI shown in (A) (scale bar: 100 μm); (C) profiles of MIP and sqrt(SV) across a microvessel in the CAM, along the axial and lateral directions.

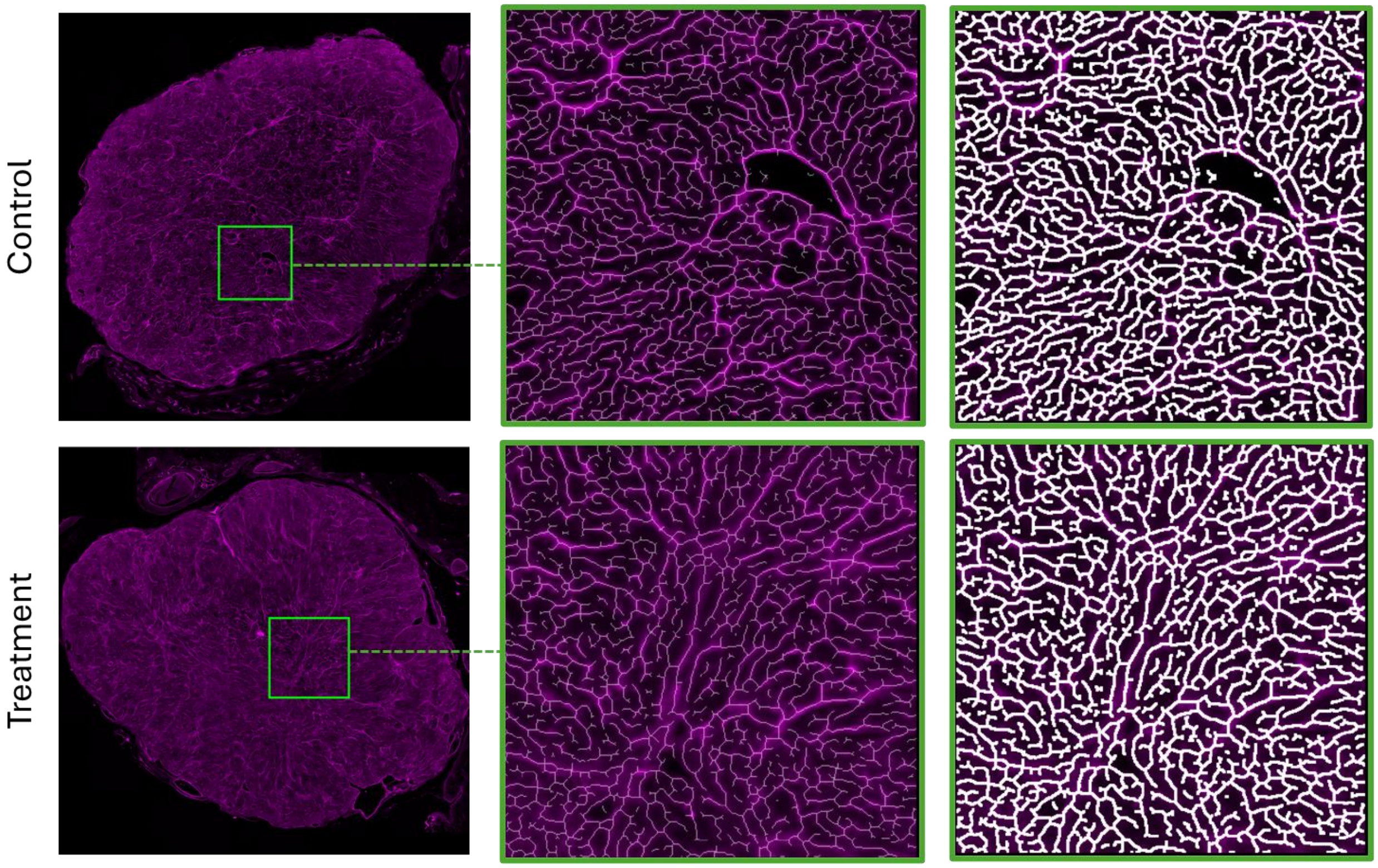

Supplementary Figure 4: Fluorescence images of histological sections of the control and treated tumors shown in Figure 9. Center sections (left, ROI = 1mm x 1mm). Zoom on the ROI (center) with overlayed skeleton. Same ROI after convolution of the skeleton with a 3x3 pixels kernel (most of the fluorescent area is covered).

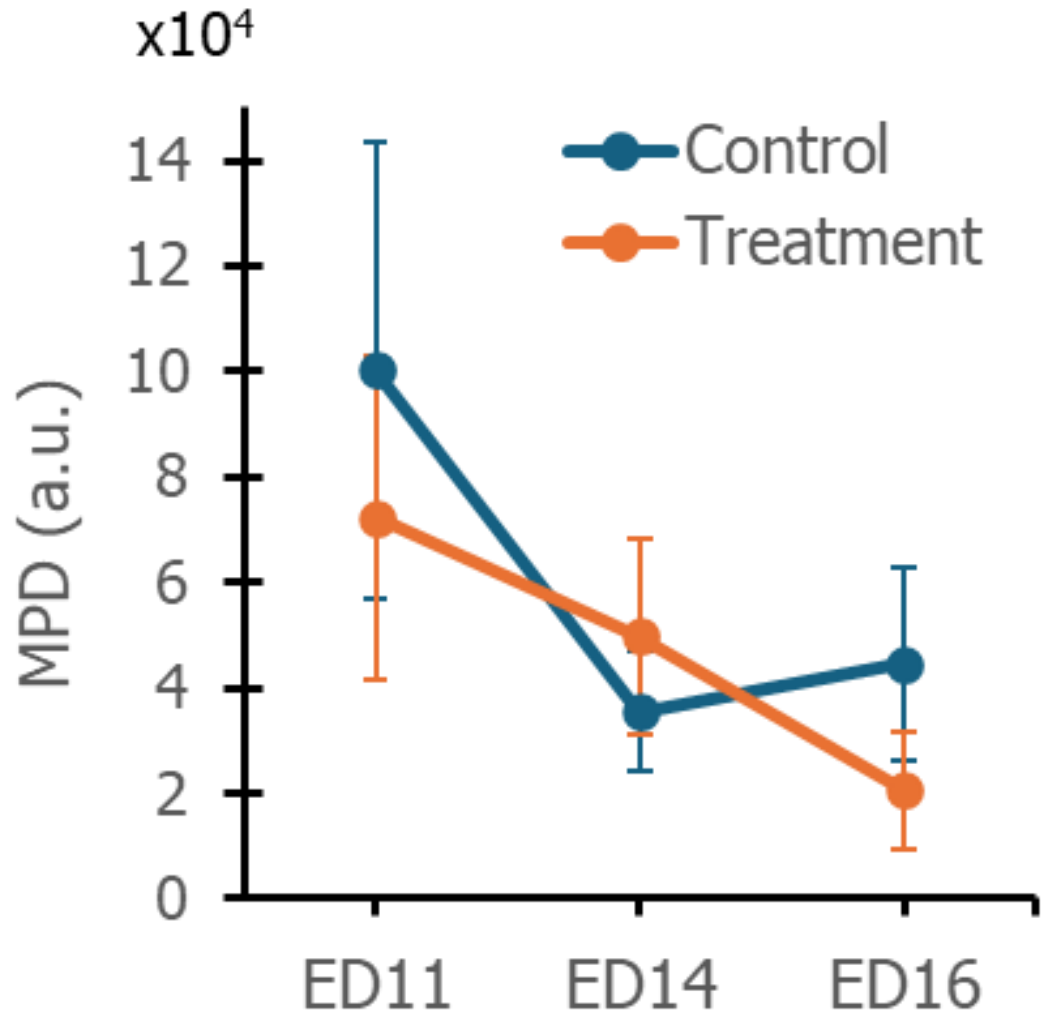

Supplementary Figure 5: Mean power Doppler measured throughout experiment timeline in control (n=10) and treated (n=11, 10 and 9 at ED11, ED14 and ED16, respectively) tumors post motion compensation and SVD filtering (auto threshold)

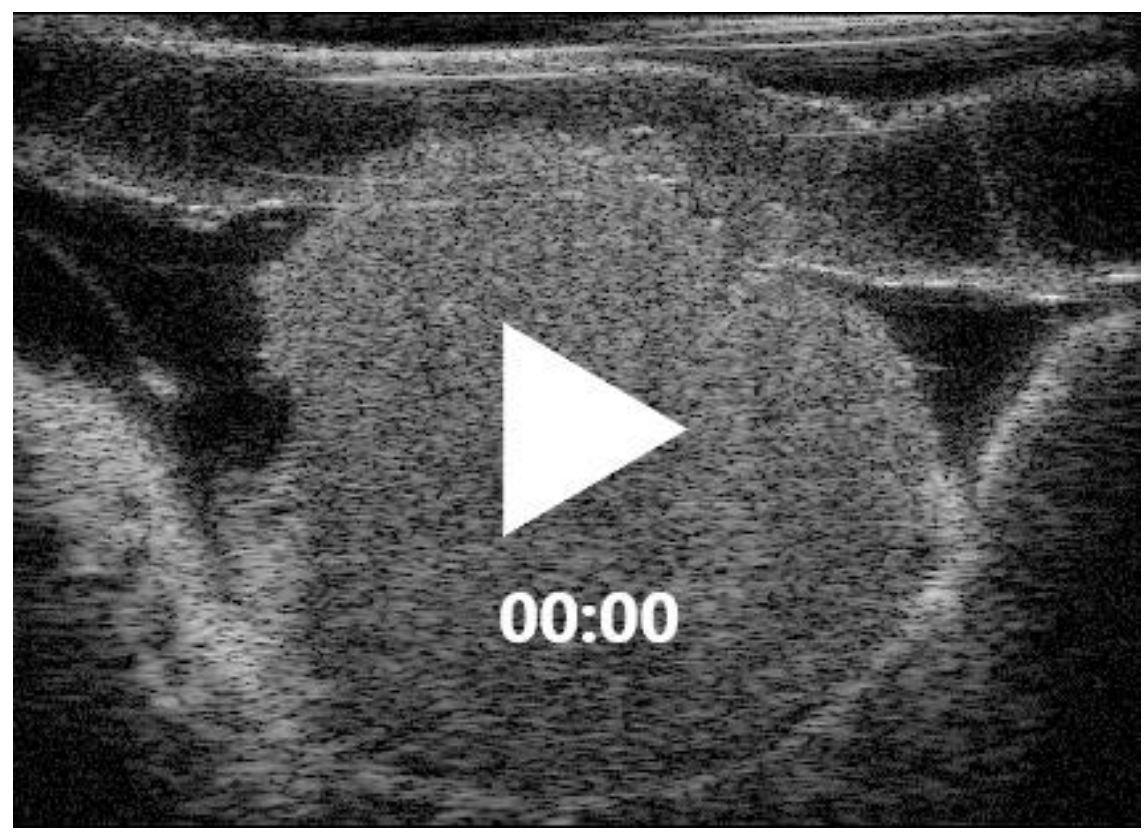

Supplementary video 1: example of a B-mode cineloop of a tumor acquired at 107 fps using a 50-MHz linear array (50dB dynamic range), after motion compensation.

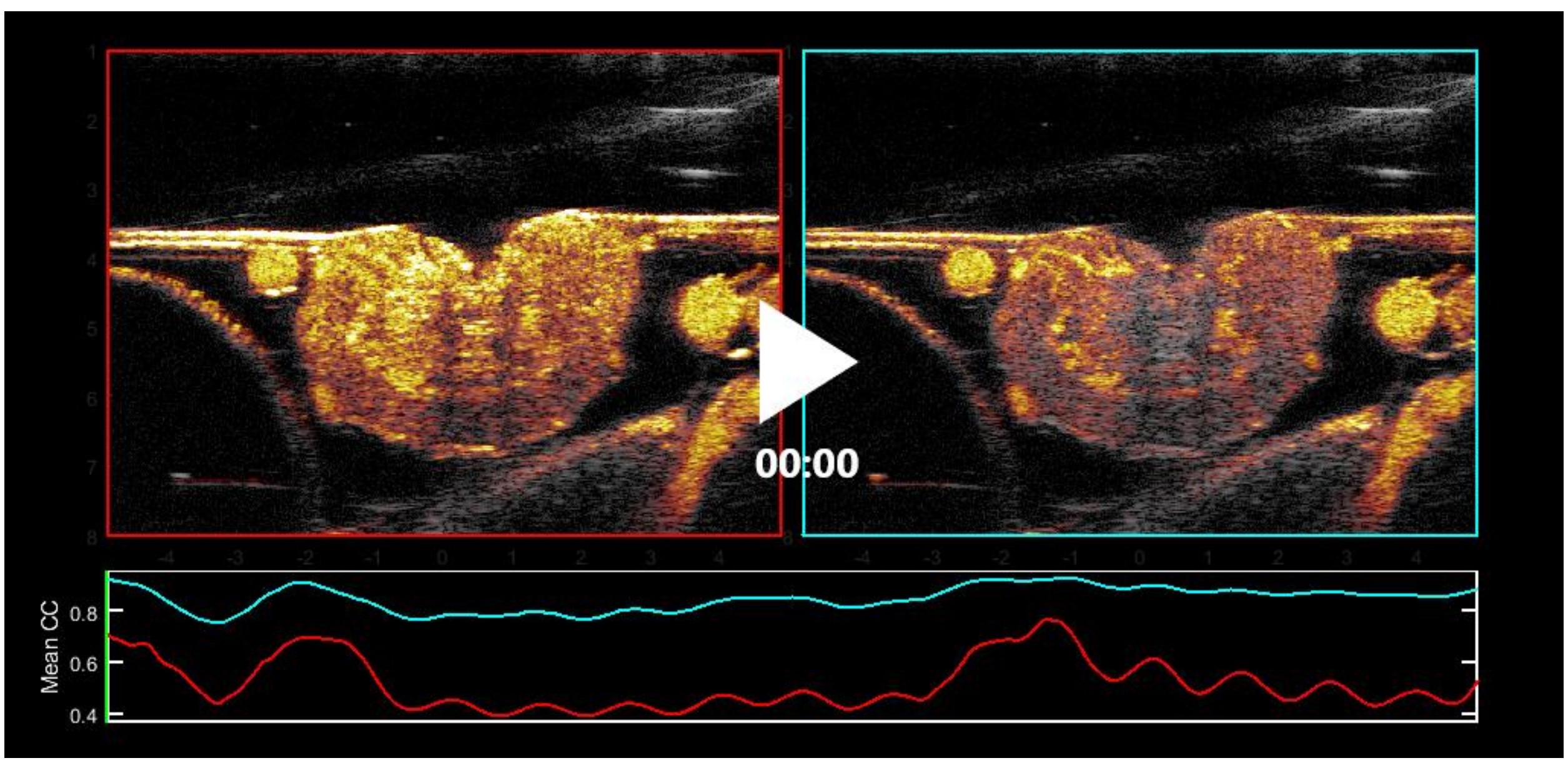

Supplementary Video 2: B-mode cineloops (grayscale) of the same CAM Renca tumor shown in Figure 7 A-H with overlaid speckle variance images calculated from short ensembles (16 frames), pre (top-left) and post (top-right) motion compensation. Comparison of the average correlation coefficients between the full frames of the ensemble and the first full frame pre and post motion compensation (bottom) demonstrates the effectiveness of the motion compensation algorithm, and its impact on tissue motion artefact affecting the speckle variance.